\documentclass[hyper,a4paper,12pt,oneside]{JHEP3}
\usepackage{amsmath,amscd}
\usepackage{amsfonts}
\usepackage{amssymb}
\usepackage{graphicx,shortvrb}
\usepackage{newlfont}
\usepackage{subfigure}

\allowdisplaybreaks[1]

\newcommand{\cno}[1]{\,\textbf{:}#1\textbf{:}\,}
\newcommand{\Jdet}{\mathrm{Jdet}}
\newcommand{\N}{{\cal N}}
\newcommand{\Op}{{\cal O}}
\newcommand{\D}{{\mathfrak D}}
\newcommand{\R}{{\mathfrak R}}
\newcommand{\J}{{\mathfrak J}}
\newcommand{\Q}{{\mathfrak Q}}
\newcommand{\K}{{\mathfrak K}}
\newcommand{\T}{{\mathfrak T}}
\renewcommand{\P}{{\mathfrak P}}
\newcommand{\dD}{{\delta\mathfrak D}}
\newcommand{\dn}{\delta n}
\newcommand{\dE}{\delta E}
\renewcommand{\H}{{\mathcal H}}
\newcommand{\F}{{\mathcal F}}
\newcommand{\W}{{\mathcal W}}
\newcommand{\spc}{~,\qquad}

\newcommand{\ab}{\mathbf{a}} \newcommand{\bb}{\mathbf{b}} \newcommand{\cb}{\mathbf{c}}

\newcommand{\sph}[1]{$\mathrm{S}^{#1}$}
\DeclareMathOperator{\SU}{\mathrm{SU}}

\DeclareMathOperator{\psu}{\mathfrak{psu}}
\DeclareMathOperator{\uu}{\mathfrak{u}}
\DeclareMathOperator{\su}{\mathfrak{su}}

\newcommand{\pd}{\partial}

\newcommand{\fdf}[2]{\frac{\delta #1}{\delta #2}}

\DeclareMathOperator{\tr}{Tr}

\newcommand{\abs}[1]{\left\vert~#1~\right\vert}

\newcommand{\ket}[1]{\left\vert~ #1 ~\right\rangle}
\newcommand{\ave}[1]{\left\langle~ #1 ~\right\rangle}

\newcommand{\bok}[3]{\Bigl\langle~ #1~\Bigl\vert~ #2~\Bigr\vert~#3~\Bigr\rangle}
\newcommand{\zbra}[1]{\Bigl\langle\relax{\kern-.4em}\Bigl\langle~#1~\Bigr\vert}
\newcommand{\zinnp}[2]{\Bigl\langle\relax{\kern-.4em}\Bigl\langle~ #1~\Bigm\vert~ #2~\Bigr\rangle}            
\newcommand{\zbok}[3]{\Bigl\langle\relax{\kern-.4em}\Bigl\langle ~#1~\Bigl\vert~ #2~\Bigr\vert~#3~\Bigr\rangle} 

\preprint{\small WIS/14/08-JULY-DPP}

\title{Weakly Renormalized Near 1/16 SUSY Fermi Liquid Operators in $\N=4$ SYM}
\author{Micha Berkooz  and Dori Reichmann  \footnote{micha.berkooz,~dor.reichmann~@~weizmann.ac.il}\\
Department of Particle Physics,\\The Weizmann Institute of Science,\\ Rehovot 76100, Israel }

\date{\today}
\abstract{We discuss a class of Fermi Liquid Operators in $\N=4$
SYM. We show that these operators are eigenstates of the full
quantum dilatation operator. We compute their 1 and 2 loop anomalous
dimensions, and show that, similar to Fermi liquids in condensed
matter systems, these corrections are suppressed by an arbitrarily
small parameter, which is the equivalent of one over the Fermi energy. These operators are, at the classical level, descendants of
1/16 BPS operators, with some scaling properties similar to those of
the 1/16 Black Holes in $AdS_5$. }
\begin{document}
\section{Introduction}\label{sec-intro}

In recent years we have witnessed considerable progress in
understanding $\N=4$ SYM away from the safe shores of the (chiral)
BPS spectrum. An important trigger to this progress was the
development of tools for perturbative calculations on both sides of
the AdS/CFT duality
\cite{Maldacena:1997re,Witten:1998qj,Gubser:1998bc}. On the CFT side
spin chains \cite{Beisert:2003yb,Beisert:2004hm}, integrability
\cite{Kazakov:2004qf,Beisert:2004ry} and unitarity techniques
\cite{Bern:1994zx} made exact results, to all orders in perturbation
theory, possible, such as the BES and FRS equations \cite{Beisert:2006ez,Freyhult:2007pz} and
BDS $n=4,5$ amplitudes \cite{Bern:2005iz}. In the AdS side the BMN
limit \cite{Berenstein:2002jq,Gubser:2002tv}, semiclassical
quantization of rotating superstrings \cite{Frolov:2002av},
integrability of the string worldsheet
\cite{Arutyunov:2004vx,Beisert:2005fw} and semiclassical solution of
strings stretching to the boundary \cite{Alday:2007hr} permit the
quantitative understanding of similar results at strong coupling. In
this paper we suggest that Fermi liquid like operators might be a
new avenue where weak and strong coupling results can be compared
(and in particular, in the context of low SUSY black holes).

A key characteristic of the recent tools mentioned above is a
proximity to unitarity bounds which reduces the complexity of the
calculation. For example the $\su(2)$ spin chain captures the
excitations of the theory near a 1/2-BPS unitarity bound
\begin{equation*}
    \Delta_1:=d -\frac12q_1-p-\frac32q_2 \geq 0~\spc
    \Delta_2:=d -\frac32q_1-p-\frac12q_2 \geq 0~,
\end{equation*}
where $d$ is the dimension of the state, $(q_1,p,q_2)$ are $\SU(4)$
charges (using Dynkin labels) and the angular moneta is set to zero.
The ground state of the spin chain has $q_1=q_2=0$, $d=p$ and
saturates both bounds. The possible excitation of this spin chain at
weak coupling (below an integer gap in dimension) are made out of a
complex scalar with weight $(q_1,p,q_2)=(1,-1,1)$. In perturbation
theory states in the $\su(2)$ spin chain can only mix within
themselves since the quantities $\Delta_1,\Delta_2$ are continuous
in $g_{ym}$. The reduced complexity is present also in the AdS side
where the corresponding states come from quantization near a string
with angular momentum in the $S^5$ (for $p$ scaling as $\sqrt N$
this is the BMN limit).

The spin chain techniques, and related ones, have proven to be of
great use in the BPS sector or close to it. As one reduces the
amount of supersymmetry, the level of complexity increases up to the
full non-SUSY case, where few precise quantitative matches exist.
In particular, the natural intermediate stage of the 1/16-BPS
spectrum, and the sector of states near the corresponding unitarity
bounds, is still out of our reach. For example, we possess some
knowledge of the sector in the free $\N=4$ theory
\cite{Kinney:2005ej} and in the classical gravity
\cite{Gutowski:2004ez} limits, however the interpolation between
these limits has seen little progress (see
\cite{Berkooz:2006wc,Kim:2006he,Janik:2007pm,Grant:2008sk} for
recent discussions).

\paragraph{\bf 1/16 Black holes:} The problem of the 1/16 operators is
particularly interesting since in the bulk these are identified as
black holes \cite{Gutowski:2004ez}. This in contrast to operators
with higher SUSY, of which they are not enough to form a black hole.
This is an indication that these black holes, although SUSY, are
sensitive to the full $N^2$ degrees of freedom of $\N=4$ SYM and can
be a useful intermediate stage to the understanding of the more
generic states of $\N=4$ SYM. Furthermore, one can argue that
similar black holes in $AdS_4$ are sensitive to the (in)famous
$N^{3/2}$ degrees of freedom in this field.

\paragraph{Fermi-sea model of the 1/16 Black holes:} In
\cite{Berkooz:2006wc} we pointed out that Fermi-sea operators in
$\N=4$ SYM may play a key role in the construction of 1/16-BPS
states in the interacting theory. The smoking gun is a scaling
between charge and angular momentum found in the black-holes
solutions (where $(q_1,p,q_2)=(0,0,q)$)
\begin{align}
    j_R/N^2\propto& \left(q/N^2\right)^{2}\,\spc\text{if}\quad j_L=j_R~\spc q\gg N^2
    \cr
    j_R/N^2\propto& \left(q/N^2\right)^{3/2}\,\spc\text{if}\quad j_L=0~\spc q\gg N^2.
    \label{bhscale}
\end{align}
These relation are typical to Fermi-sea models where each fermion carries charge $q=1$ and the levels are graded by the angular momentum $j_R$. Filling the levels up to some Fermi-level K we find a relation
\begin{equation}
    (j_R/N^2)\propto (q/N^2)^{(m+1)/m}~, \end{equation}
with $m$ is the dimensionality of the Fermi-sea for, and $N^2$ is
the degeneracy of fermions. $j_L=j_R$ is naturally associated with
$m=1$ and $j_L=0$ is naturally associated with $m=2$.

Even though the scaling is similar, the precise coefficients in front of the relations are different for the Fermi sea operators and the Black holes. We are therefore still missing some components in the operators that correspond to the black holes, but we can still ask what are the algebraic properties of the Fermi sea by themselves, as these properties may carry over to the full operators that correspond to the black holes. 

In particular, the operators here are not primaries, and they are not precisely the same operators discussed in \cite{Berkooz:2006wc}, where an attempt to find BPS primaries was carried out. However, they are very similar to the latter in the sense that we can go to the operators of \cite{Berkooz:2006wc} by adding an additional insertion of the gauge field strength (and additional insertions of scalars to go to the general case discussed there). Without this insertion, which is the case here, the operators cannot be BPS primaries as they does not satisfy the unitarity bound ($d=2+2j_L+\frac12q_1+p+\frac323q_2$), and indeeds the operators are descendants of non-BPS operators. 

We would also like to emphasize that the results of \cite{Berkooz:2006wc} were limited to tree-level (with weak coupling). The tools discussed in this work can be used to refine the calculations there to include 1-loop contributions. In \cite{Berkooz:2006wc} we argued that certain operators cannot be written as $Q$ of something, but we did not identify which combination was primary. In order to do so, and verify or refute the construction there using the tools in the current paper, one still needs to solve an operator mixing problem. However we choose to postpone this investigation to future work and rather take a new direction.

\paragraph{Fermi-sea and Fermi liquids:} The Fermi-sea operators/states are
created by fermions belonging to the $\su(1,1)$ sector of $\N=4$
SYM. In principle they are similar to Fermi-sea states common in
condensed matter physics, however our fermions come in an
$N^2-1$-fold degeneracy, arising from the $\SU(N)$ gauge group.
Another peculiarity of our Fermi-sea is that it carries large
angular momentum, and therefore states are not graded and filled
only by energy, which leads to a filling which is the isotropic, but rather the momentum shells which are filled are
confined to 1-dim "lightcone" in momentum space (in the $j_L=j_R$
case, which will be our main focus). Despite the strange nature of
the fermions we will demonstrate that they form a normal Fermi
liquid just the same way interacting fermions do in earthly physics.

In a sense, the appearance of Fermi-sea operators in the AdS/CFT
correspondence could be expected, since they are the ground state of
choice of many condensed matter system. Let us review briefly the
notion of normal Fermi liquid. The phenomenological theory of Landau \cite{Landau:1956aa}
\footnote{See \cite{Nozires:1966aa} for a comprehensive review.} describes the state of strongly coupled
fermions at low temperature. This theory, originally constructed for
liquid ${}^3\mathrm{He}$, was found to be a good description
(generalized for charged fermions) in much more general situations
ranging from electron in metals (when not undergoing a
superconductor phase transition, such as Alkali metals) to highly
dense matter in stars.

Landau \cite{Landau:1956aa} considered a non-interacting gas of
fermions at equilibrium. At high temperatures $T$ the distribution
of particles with energy $\epsilon$ is given by the well known
Fermi-Dirac distribution. In this regime, the kinetic term
usually dominant, and any interaction between the fermions is
treated as a perturbation. As the temperature is lowered,
$T\rightarrow 0$, the Fermi-Dirac distribution approaches a step
function at the Fermi-energy (fixed by the chemical potential). At
the same time a critical temperature appears when interactions
between the fermions become important.
Fermion systems
where the particle interaction and the exclusion principle act
simultaneously are often named degenerate Fermi liquids.

The degenerate Fermi liquid at $T=0$ is described by a Fermi-sea; it's elementary excitations are quasiparticles and
quasiholes in analogy to the non-interacting case. However the
nature of the quasiparticles could be very different from the
excitation of the non-interacting system. Landau's phenomenological
model describes the energy of a the degenerate Fermi liquid in terms
of an the quasi-particle density, the effective Fermi velocity
(effective mass) and effective couplings (known as Landau's
parameters)
\begin{equation}\label{ELFLM}
    E = \sum_k v_F^*(\abs{k}-k_F)\delta n(k)+\frac12\sum_{k,k'}f^*(\abs{k-k'})\delta n(k)\delta n(k')~, \end{equation}
with $\delta n(k)$, is the density of quasi-particles, exhibiting a
Fermi-Dirac distribution in equilibrium (with zero chemical
potential). In principle, the various couplings can be computed by
flowing from the UV, but, as expected, this is usually difficult in
practice. However, it is often the case that the excitation of
quasi-particles near the Fermi-surface are long-lived \footnote{Due
to the interaction any quasiparticles, above the Fermi surface, can
decay into two quasiparticles closer to the Fermi surface, however
the decay rate behaves as $1\big/{(E_{q.p}-E_F)^2}$.} allowing to
treat them as stable excitation and allowing for a perturbative
expansion in $T/E_f$. It is quite remarkable that this simple and
intuitive model, and the introduction of the $T/E_f$ perturbation
theory, are strong enough to describe all the observed phenomena of
${}^3\mathrm{He}$, and other normal Fermi liquids.

The main obstruction for the appearance of a normal Fermi liquid
is a breaking of symmetry. The most famous examples are the BCS
mechanism (where a phonon-electron interaction drives a condensation
of Copper pairs) and the Wigner crystal (where the translation
symmetry breaks at low densities).

\medskip

In this work we analyze the fate of the Fermi liquid operators of
$\N=4$ SYM in perturbation theory. We show that \begin{itemize}
\item Based on symmetries, we prove that the Fermi-sea state is an eigenvalue of
the dilatation operators to all order in perturbation theory and
does not mix with any other operator in $\N=4$ SYM.
\item We compute its 1 and 2 loop anomalous dimension and find a non-trivial
cancelation at weak coupling leading to a much milder anomalous
dimension compared to the naive expectation from familiar results of
spin chain states. The corrections to the classical dimension are
suppressed by one over the analogue of the Fermi energy, suggesting
that the operators can be traced from weak to strong coupling.
\item Investigating quasi-particles energies near
the Fermi-sea, we find that the cancelation allows us to write a
consistent Landau Fermi liquid model \eqref{ELFLM} with a well
defined $1/E_f$ expansion.
\end{itemize}

The $1/E_f$ expansion sheds new light on the problem of 1/16 BPS operators. One route is to continue the search for a large degeneracy of operators, which remain 1/16 BPS when field theoretic loop correction are taken into account. However a new scenario is possible which suggests that SUSY and the degeneracy are lifted by $\alpha'$ or stringy loop corrections.

The paper is organized as follows. In section \ref{sec-pre} we recap some notations and tools for $\N=4$ SYM . In section \ref{sec-gs} we discuss the fermionic $\su(1,1)$ sector of $\N=4$ and the 1-dim Fermi-sea operators within weak coupling perturbation theory. In section \ref{sec-qp} we show that indeed the Fermi-sea operator is a ground state of a Fermi liquid. In section \ref{sec-2d} we briefly discuss a possible generalization of the work to a 2-dim Fermi-sea. We conclude in section \ref{sec-end} with some additional discussion and
directions for future research.

\section{Tools for $\N=4$ super Yang-Mills theory}\label{sec-pre}

In this section we briefly recall some features of
$\N=4$ SYM and it's $\psu(2,2|4)$ algebra, which we will need later
on. We follow the notation of \cite{Beisert:2004ry} - readers
familiar with this work may skip this section.

The component of the $\N=4$ multiplet are
\begin{itemize}
  \item The gauge field strength $F_{\alpha\beta}$ and $\bar F_{\dot\alpha\dot\beta}$.
  \item The gauginos $\Psi_{\alpha a}$ and $\bar\Psi_{\dot\alpha}^a$.
  \item The complex scalars $\Phi_{ab}$ with the antisymmetry $\Phi_{ab}=-\Phi_{ba}$~. \end{itemize}
The undotted Greek letters ($\alpha,\beta,\ldots$), dotted Greek
letters ($\dot\alpha,\dot\beta,\ldots$) and Latin letters
($a,b\ldots$) stands for $\SU(2)_L$, $\SU(2)_R$ and $\SU(4)$
fundamental indices. Raising and lowering the $\SU(4)$ indices
changes between the fundamental and anti-fundamental
representations. The scalars belong to a $\mathbf{6}$ of $\SU(4)$
and obey the reality condition
\begin{equation}
    (\Phi_{ab})^\dag  := \bar \Phi^{ab} = \frac12\epsilon^{\,abcd}\Phi_{cd}~.
\end{equation}
The gauge group is $G=\SU(N)$, and all fields transform in the
adjoint representation. When we will need to be specific about the
gauge group structure we will write all fields as $\W=\W^at^a$ with
$a=1,\ldots \dim G$, and $t^a$ are generators of $\SU(N)$
\footnote{We use the same letters for gauge group and $\SU(4)$
indices, it will be clear to distinguish between them from the
context.}. The covariant derivative is
\begin{equation}
    D_{\alpha\dot\alpha}\W:=(\sigma^\mu)_{\alpha\dot\alpha}\left(\pd_\mu\W-i[A_\mu,\W]\right)~.
\end{equation}
We use the fundamental representation for $\SU(N)$ with the following properties
\begin{equation}
    \tr t^at^b = \delta^{ab}\spc
    \sum_a t^at^a = \frac{N^2-1}{N}\mathbb{I}~.
\end{equation}
We use the coupling constat $g^2$ proportional to the 't~Hooft coupling
\begin{equation}
    g^2 = \frac{g_{ym}^2N}{8\pi^2}~.
\end{equation}

\paragraph{The oscillator representation and spin chains - }In the planar limit a useful set of states are spin chains. These are states which are mapped to single trace local operators
\begin{equation}
    \ket{\W_1\W_2\ldots \W_n} \mapsto \tr (\W_1(x)\W_2(x)\ldots \W_n(x))~. \end{equation}
For spin chains all the gauge group information is encoded in the
location inside the chain (relative to other spins). This allows one
to write each spin using the oscillator representation. There are
three types of oscillators - left-handed bosons, right-handed bosons
and $\SU(4)$ charged fermions - which satisfy:
\begin{equation}\label{oscilator}
    \left[\ab^\alpha,\ab^\dag_{\beta}\right]=\delta^\alpha_\beta\spc
    \left[\bb^{\dot\alpha},\bb^\dag_{\dot\beta}\right]=\delta^{\dot\alpha}_{\dot\beta}\spc
    \left\{\cb^a,\cb^\dag_b\right\}=\delta^a_b~.
\end{equation}
Using these oscillators one can build all possible spins
\begin{align}\label{parton-full}
    D^k F :=& (\ab^\dag)^{k+2}(\bb^\dag)^{k}~~~(\cb^\dag)^0\ket{0} \cr
    D^k \Psi :=& (\ab^\dag)^{k+1}(\bb^\dag)^{k}~~~(\cb^\dag)^1\ket{0} \cr
    D^k \Phi :=& (\ab^\dag)^{k}~~~(\bb^\dag)^{k}~~~(\cb^\dag)^2\ket{0} \cr
    D^k \bar\Psi :=& (\ab^\dag)^{k}~~~(\bb^\dag)^{k+1}(\cb^\dag)^3\ket{0} \cr
    D^k \bar F :=&
    (\ab^\dag)^{k}~~~(\bb^\dag)^{k+2}(\cb^\dag)^4\ket{0}~. \end{align}
The $\psu(2,2|4)$ algebra of the zero coupling theory is made out of
bilinears of oscillators (see appendix \ref{app-par} for details).
The generator of $\psu(2,2|4)$ are expected to receive quantum
correction, it is possible to choose a regularization scheme such that
operator mixing occurs only between operators with the same
zero-coupling dilatation (dimension). Using these scheme
all operators can be written as a formal expansion in $g$
\begin{equation}
    \mathcal{X}(g)=\sum_{n=0}^\infty\mathcal{X}_n g^{n}
    \spc
    \left[\D_0,\mathcal{X}_n\right]=d_{\mathcal{X}}\mathcal{X}_n
    \spc
    \mathcal{X}\in\psu(2,2|4)~. \end{equation}
All these operators depend of course on $N$ - the rank of the gauge
group - but we will suppress this dependence in the notation.

The anomalous dimension operator $\dD:=\D-\D_0$ commutes with all the
$\psu(2,2|4)$ algebra. It has an expansion
\begin{equation}
    \dD = \sum_{n=2}^\infty\dD_n g^{n}~.
\end{equation}
For spin-chains in the planar limit the leading term of the
anomalous dimension is given by the 'Harmonic action'
\begin{equation}
    \dD_2 = \sum_{k=0}^{L}H_{k,k+1}\spc
    H_{1,2}\ket{AB}=\sum_{C,D}c_{n,n_{12},n_{21}}\ket{CD}~,
\end{equation}
where $L$ the length of the spin chain, and $k$ runs over the spin
location in the chain. The two-spin operator $H_{1,2}$ is evaluated
by considering all possible two spin states generated by having an
oscillator hop from one site to the other. The coefficient
$c_{n,n_{12},n_{21}}$ is a function of the total number of
oscillator in the two spins (conserved by $H_{1,2}$), $n_{12}$ the
number of spins jumping form site 1 to 2 and $n_{21}$ the number of
spins jumping from site 2 to 1
\begin{equation}
    c_{n,n_{12},n_{21}}=\begin{cases}
    (-)^{1+n_{12}n_{21}}\frac{\Gamma\left(\frac12\left(n_{12}+n_{21}\right)\right)\Gamma\left(1+\frac12\left(n-n_{12}-n_{21}\right)\right)}{\Gamma\left(1+\frac12n\right)} &n_{12}+n_{21}>0 \\
    h(n)     & n_{12}=n_{21}=0
    \end{cases}
\end{equation}
where $h(n)$ are the harmonic numbers
\begin{equation*}
    h(n):=\sum_{k=1}^{n}\frac1{k}~,\qquad h(0):=0~. \end{equation*}

\paragraph{The anomalous dimension operator and closed sectors:}
The planar expression can be lifted to the full theory at finite
$N$. Using Feynman diagrams it can be proven that the leading term
(in $g^2$) in $\dD$ has the general form
\begin{equation}\label{spin-finite1}
    \dD_2 = -\frac1{N}\sum_{A,B,C,D}\mathcal{C}^{AB}_{CD}\tr\cno{\left[\W_A,\check \W^C\right]\left[\W_B,\check \W^D\right]}~,
\end{equation}
where the summation is over all partons in the theory \footnote{i.e.,
elementary fields acted by covariant derivatives, $\mathcal W =
\begin{Bmatrix} D^{k}F, & D^{k}\Psi_i, & D^{k}\Phi_{ij}, &
D^{k}\bar\Psi^i, & D^{k}\bar F
\end{Bmatrix}$, with all lorentz indices symmetrized.}. The symbol $\W_A$
stands for the parton and $\check\W^A$ are functional derivatives
with respect to the partons
\begin{equation}
    (\check\W^A)^a = \fdf{}{(\W_A)^a}\spc
    a=1,2\ldots\dim G~. \end{equation}
The coefficient $\mathcal{C}^{AB}_{CD}$ are calculated directly from
the Feynman diagrams.
Computing $\dD_2$ on the spin chain determines it completely.
Indeed, in the full theory
\begin{equation}\label{spin-finite2}
    \mathcal{C}^{AB}_{CD} = \frac12(-)^{\zeta_C(\zeta_B+\zeta_C)}\bok{AB}{H_{12}}{CD}~,
\end{equation}
where the LHS is the full expression and the RHS is the planar
expression ($\zeta_X$ is the fermion number, $\zeta_X=1(0)$ for
fermion(boson)). This is just the statement that at order $g^2$, all Feynman diagrams contributing to the anomalous dimension are planar. However for higher correction there are non-planar diagrams which are not captured by
the spin chains.

In general the anomalous dimension operator mixes different
operators (states). However one can use a regularization scheme where the
Poincar\'e group and R-symmetry do not receive quantum corrections.
In this scheme operator which have a definite charge under
$\SU(2)_R\times\SU(2)_L\times\SU(4)\times\mathrm{U}(1)_{\D_0}$ can
mix only with operators with the same charges. A sector of the
theory is a set of states closed under mixing, i.e cannot mix with states outside the
sector, due to conservation of the above charges. An example is the
$\mathfrak{sl}(2)$ sector which is generated by a single scalar and
lightcone derivatives:
\begin{equation}
    \W\in\left\{\Phi_{34},~D_{1\dot1}\Phi_{34},~(D_{1\dot1})^2\Phi_{34},~\ldots\right\}~.
\end{equation}
All states in the sector has charges $(d_0 ; j_L,j_R ; q_1,p,q_2 ;
L)=(s+p ; s,s ; 0,p,0 ; p)$. To prove that there are no other states
in the theory with these charges one uses the fact that all parton
of the theory obey $d_0\geq j_L+j_R$. Among the partons with $d_0=
j_L+j_R$ only $(D_{1\dot1})^n\Phi_{34}$ has $q_1+q_2=0$. The full
set of closed sector and many more details may be found in
\cite{Beisert:2004ry,Beisert:2003jj}.

\section{The Fermi liquid ground state}\label{sec-gs}

The Hilbert space of zero coupling $\N=4$ SYM on \sph3 is the Fock
space generated by the mode expansion of the elementary fields on
\sph3, with the additional constraint of gauge invariance. Applying
the operator-state mapping, these states correspond to local gauge
invariant operators generated by the partons $\pd^{k}F,
\pd^{k}\Psi_i,\pd^{k}\Phi_{ij},\pd^{k}\bar\Psi^i,$ and $\pd^{k}\bar
F$. The Hamiltonian of the theory on \sph3 is mapped to the
dilatation operator, which in the notation discussed before and in
$g=0$ is
\begin{equation}
    \H = \D = \sum_{A}d_0\tr\cno{\W_A\check \W^A}~, \end{equation}
where $d_0$ is the classical dimension of $\W_A$. We often switch
back and forth between energy of a state and the dilatation charge
of a local operator, not to be confused with the energy in Minkowski
space.

In a grand canonical ensemble the distribution functions is governed
by the Gibbs free energy, which is just the Hamiltonian shifted by
chemical potentials for a set of $\mathrm{U}(1)$ charges. The charges include
all global internal symmetries as well as the angular momenta on the
\sph3
\begin{equation}
    \beta\F = \beta\H-\sum_{I=1}^5\mu_IQ_I~, \end{equation}
where $\beta$ is the inverse temperature, the $\mu_i$ are the
chemical potentials and $Q_I$ are the corresponding charge operators.
The ground state is the state (or a set of states) minimizing the
free energy at zero temperature. We will consider twisted sector of
the theory where $\mu_I\propto\beta$ (cf. \cite{Harmark:2007px}).

The Fermi-sea operators appear in the ensemble for a specific choice of chemical potentials, such that $\F$,
localizes near a $\frac1{16}$-BPS bound, i.e
\begin{equation}
    \F = \left(\D-2\J_R^{\,3}-\frac12\mathfrak{q}_1-\mathfrak p-\frac32\mathfrak q_2\right)+O(\beta^{-1})~,
\end{equation}
where $\J_R^{\,3}$ is the right-handed angular momentum charge and
$(\mathfrak q_1,\mathfrak p,\mathfrak q_2)$ are $\SU(4)$ charges
(see appendix-\ref{app-par} for the definition of these charges).
For the specific choice above, all partons operators have a
non-negative contribution to the free energy. The ground state in
this sector is generated by partons which has vanishing contribution
\begin{equation}\label{sector-large}
    \pd_{\alpha_1\dot 1}\pd_{\alpha_2\dot 1}\cdots \pd_{\alpha_n\dot 1}\mathcal{X}
    ,\qquad
    \mathcal{X}\in\left\{\Psi_{\alpha_{n+1}4}~,~ \bar F_{\dot1\dot1}~,~ \Phi_{i4}~,~ \bar\Psi^{i}_{\dot
    1}~|~i=1,2,3\right\}~. \end{equation}

Turning on weak Yang-Mills coupling generates anomalous dimension
and mixing between the local operators. However, since the zero
coupling dilatation charge is conserved by Feynman diagrams, the
difference $d_0-2j_R^3-\frac12q_1-p-\frac32q_2$ is also conserved
and operators generated from \eqref{sector-large} \footnote{Derivatives are replaced by covariant derivatives} can mix
only within themselves (due to the integer gap when inserting
operators outside this sector). As a consequence the Hilbert space
generated by the partons of \eqref{sector-large} forms a closed
sector of $\N=4$ SYM. The partons of this sector form a
representation of $\su(1,2|3)$ algebra, which is common to use as a
name for the sector \cite{Beisert:2004ry}.

The anomalous dimensions, or mixing, do not vanish for all operator
in the sector. Rather it is encoded in
\begin{equation*}
    \F_{(\su(1,2|3))} = \left(\D-2\J_R^{\,3}-\frac12\mathfrak q_1-\mathfrak p-\frac32\mathfrak q_2\right) = (\D-\D_0)=\dD~. \end{equation*}
An operator is a 1/16 BPS, at finite values of $g_{ym}$, iff
anomalous dimension vanishes ($\dD=0$). Such ground states of the
$\su(1,2|3)$ sector are part of semi-short multiplets (see
\cite{Grant:2008sk} for further discussion). Classifying all exact
1/16-BPS operators is a very interesting open problem.

 In the following we will treat the
anomalous dimension operator $\dD$ as the Hamiltonian. The operator
$\dD$ will account for both weak coupling effects, i.e mixing and
anomalous dimension. The Fermi liquid appears when we study
eigenstates of this Hamiltonian.

The 1-dim Fermi-sea comes about when one discusses the maximal $j_L$
representation allowed by unitarity \cite{Dolan:2002zh}, i.e., the first
line in \eqref{bhscale}. In the parton picture one finds it by
restricting the theory to the sector known as the fermionic
$\su(1,1)$ sector \cite{Beisert:2004ry,Beisert:2003jj}. The
restriction is done by using only partons such that
\begin{equation}
    (q_1,p,q_2) = (0,0,q_2)~,
    \qquad j_L=j_R+\frac{q_2}{2}~. \end{equation}
The only partons obeying this are
\begin{equation}\label{parton-su11}
    \psi_{(n)} := \frac1{(n+1)!}\left(D_{1\dot1}\right)^n\Psi_{1\,4}~, \end{equation}
where the normalization is chosen for convinces in later
calculations. We will refer to the number of derivatives as the
level ($n$), and the angular momentum (to which we will loosely
refer as momentum from now on) is $((n+1)/2,n/2)$.

Operators with some scaling properties similar to the 1/16 BH -
 i.e. equation \eqref{bhscale} - are constructed as follows.
Consider a state in this sector with charge $q_2$ and minimal
conformal dimension. The charge dictates that we use $q_2$ fermionic
operators of type \eqref{parton-su11} and for minimal dimension it
best to have as little amount of derivatives as possible. However,
as the fermions obey Pauli's exclusion principle
we can use each fermion only once (bearing in mind that the quantum
numbers of the fermions also include an adjoint gauge index). The
1-dim Fermi-sea operator is then
\begin{equation}\label{Fermi-sea-1dim}
    \Op^{(K)}_{\mathrm{1dim}}
    :=\prod_{n=0}^{K}\psi_{(n)}^1\psi_{(n)}^2\cdots\psi_{(n)}^{\dim G}
    :=\prod_{n=0}^{K}\Jdet\left(\psi_{(n)}\right)~. \end{equation}
The superscripts are the adjoint representation indices, the
subscripts encode the level of the parton and in the final equation
we introduced the notation $\Jdet$ to denotes the product of the all
the fermions at the same level. Note that each $\Jdet$ is gauge
invariant on its own. In effect it is the volume form of the group.
All fermionic operator are evaluated at the same space point, correspondingly the
expression can be viewed as a state in radial quantization.

The 1-dim Fermi-sea at zero coupling, large $N$ and large $K$ has
the following charge, dimension and angular momenta:
\begin{subequations}
\begin{align}
\label{1dimch}
    (q_1,p,q_2)=&\,\sum_{n=0}^{K}\sum_{a=1}^{\dim G}\left(0,0,1\right)=(N^2-1)(K+1)\left(0,0,1\right)
    \approx\,\left(0,0,N^2K\right),
\\
\label{1dimdim}
    d_0 =\,&\sum_{n=0}^{K}\sum_{a=1}^{\dim G}\left(\frac32+n\right)=(N^2-1)\frac{(K+3)(K+1)}{2}
    \approx\,\frac{N^2K^2}{2},
\\
\label{1dimang}
    \left(j_L,j_R\right)=&\,\sum_{n=0}^{K}\sum_{a=1}^{\dim G}\left(\frac{n+1}{2},\frac{n}{2}\right)=(N^2-1)\frac{K(K+1)}{4}\left(\frac{K+2}{K},1\right)\cr
    \approx&\,\left(\frac{N^2K^2}{4},\frac{N^2K^2}{4}\right)~. \end{align}
\end{subequations}

Eliminating $K$ from equation \eqref{1dimch} and inserting it into equation
\eqref{1dimang}, we obtain the scaling relation \eqref{bhscale}. Note, however,
that the coefficient in front of this relation (omitted in
\eqref{bhscale}) does not come out correctly, nor have we identified
a large entropy ensemble of operators.

This operator is rather unique in that it does not mix with any
other operator in perturbation theory. This is easy to see
even without considering the explicit form of the quantum corrected
dilatation operator. To do so we consider the effect of weak
coupling on \eqref{Fermi-sea-1dim}, and check all possible mixing. It is enough to check with which operators
$\Op^{(K)}_{\mathrm{1-dim}}$ may have a non-vanishing correlator.
Since the 1-dim Fermi-sea belongs to the fermionic $\su(1,1)$ sector,
it can only mix with operators in this sector. Due to conservation
of the $q_2$ charge the only possibility is a momentum hoping, i.e
\begin{equation*}
    \left(\cdots D_{1\dot1}\psi_{1\;4}\right)\left(\cdots\psi_{1\;4}\right)
    \leftrightarrow
    \left(\cdots\psi_{1\;4}\right)\left(\cdots D_{1\dot1}\psi_{1\;4}\right)~.
\end{equation*}
However since angular momentum is conserved, the state must be
identical to the original, or there will be two fermion with same
momentum which is forbidden by Pauli's exclusion principle. Thus the
1-dim Fermi-sea operator is an eigenstates of $\dD$. In the
following subsection we will examine this statement in one and two
loops.

\subsection{The $g^2$ Hamiltonian}\label{sec-gs-1loop}

There are a few ways to derive the $g^2$ Hamiltonian - Feynman
diagrams, the harmonic action (discussed in section \ref{sec-pre})
and algebraic structure. We will use the latter algebraic method
since it is easiest to extend it to the $g^4$ Hamiltonian.

The algebraic method was first introduce by Beisert
\cite{Beisert:2004ry,Beisert:2003jj}. The first step is to address
the operation of the $\psu(2,2|4)$ on the fermionic $\su(1,1)$
sector \footnote{For order $g^4$ we will need to consider the
$\psu(1,1|2)$ sector, but at the order $g^2$ we simply quote the
results already at the fermionic $\su(1,1)$ sector.}. This sector is
preserved by an $\su(1,1)\times\uu(1|1)$  subgroup of $\psu(2,2|4)$,
which acts as follows:
\begin{itemize}
\begin{subequations}
  \item The $\su(1,1)$ algebra is generated by
\begin{align}
    &\J^0(g) = -\mathcal{L}+2\D_0+\dD(g)&
    &\J^{++}(g) = \P_{1\dot1}&
    &\J^{--}(g) = \K^{\dot11}(g)~,
\end{align}
where $\mathcal{L}$ is the length (parton number) operator.
$\mathcal{L}$ commutes with the entire $\su(1,1)$ algebra.
  \item The $\uu(1|1)$ algebra is generated by
\begin{align}
    &\T^+(g)=\bar\Q_{\dot 2\;4}(g),&
    &\mathcal{L}&
    \cr
    &\bar\T^-(g)=\bar{\mathfrak{S}}^{\dot 2\;4}(g),&
    &\dD(g)=2\left\{\T^+(g),\bar\T^-(g)\right\}~.&
\end{align}
\end{subequations}
\end{itemize}
In the above expressions we made explicit use of the relation
between the charges in the sector, for example
~$\D_0=2\mathfrak{L}_1^1-\frac12\mathcal{L}=2\mathfrak{\bar
L}_{\dot1}^{\dot1}-\frac32\mathcal{L}$. Also, for these relations to
work it is important to use a regularization scheme (when computing
Feynman diagrams) such that the momentum ($\P$), the lorentz
rotations ($\mathfrak{L}~,~\mathfrak{\bar L}$) and R-symmetry
($\mathfrak{R}$) receive no quantum correction.

While the $\su(1,1)$ algebra acts on the $\su(1,1)$ sector partons at
zero coupling, the action of the $\uu(1|1)$ algebra vanishes at
$g^0$, except $\mathcal{L}$. Next we write the generators as
\begin{equation}
    \mathfrak{X}=\sum_{n=0}^\infty \mathfrak{X}_n\,g^n~,\qquad \mathfrak{X}\in\su(1,1)\times\uu(1|1)
\end{equation}
where $g^2:=\frac{g_{ym}^2N^2}{8\pi^2}$ is the 't~Hooft coupling. The
generator $\bar\T^\pm(g)$ may have a $g^1$ correction, forcing the
first correction to $\dD$ being at $g^2$. The $g^1$ terms in
$\su(1|1)$ are determined algebraically as follows: The
product structure of $\su(1,1)\times \uu(1|1)$ remains, leading to
the following commutators
\begin{align}
    &\left[\J^{++}_0,\T^+_1\right]=0 &
    &\left[\J^{--}_0,\T^+_1\right]=0&
    &\left[\J^3_0,\T^+_1\right]=0 &
    \cr
    &\left[\J^{++}_0,\bar \T^-_1\right]=0 &
    &\left[\J^{--}_0,\bar \T^-_1\right]=0&
    &\left[\J^3_0,\bar \T^-_1\right]=0&~.
    \label{cmtldng}
\end{align}
Considering a spin chain state, the leading order representation of the algebra is
\begin{align}
    &\J^{++}_0\ket{\psi_{(n)}} = (n+2) \ket{\psi_{(n+1)}}&
    &\J^{--}_0\ket{\psi_{(n)}} = n \ket{\psi_{(n)}}&~.
    \label{jleading}
\end{align}
From the Feynman diagrams expansion (and charges) we know that at order $g^1$ the generator $\T^+$ changes a single spin state to a two spin state while $\bar\T^-$ does the opposite. Thus the most general action for the generators consistent with the conserved charges is
\begin{subequations}
\begin{align}
    \T^+_1\ket{\psi_{(n)}}=&\,\sum_{m=0}^{n-1}a_{n;m}\ket{\psi_{(m)}\psi_{(n-1-m)}}
\\
    \bar \T^-_1\ket{\psi_{(m)}\psi_{(n)}}=&\,\bar a_{n;m}\ket{\psi_{n+m+1}}~.
\end{align}
\end{subequations}
Inserting this equation and \eqref{jleading} into \eqref{cmtldng}
allows us to determine $a_{m,n}$ and ${\bar a}_{m,n}$ up to an
overall factor (which is calculated from comparison to Feynman
diagrams). One subtlety in the calculation is that the following
state is identified with zero
\begin{equation}
    \ket{\psi_{(0)}\psi_{(n)}}+\ket{\psi_{(n)}\psi_{(0)}}\widehat{=}\, 0~,
\end{equation} since it is a gauge variation of another expression, and hence will vanish in
all gauge invariant expression. This identification is crucial for
the existence of a non-trivial solution for the generators. The final result is
\begin{subequations}
\begin{align}
    \label{1loop+}
    \T^+_1\ket{\psi_{(n)}}=&\,\frac1{\sqrt2}\sum_{m=0}^{n-1}\ket{\psi_{(m)}\psi_{(n-1-m)}}
\\
    \label{1loop-}
    \bar\T^-_1\ket{\psi_{(m)}\psi_{(n)}}=&\,\frac1{\sqrt2}\left(\frac1{n+1}+\frac1{m+1}\right)\ket{\psi_{n+m+1}}
\end{align}
\end{subequations}
Here we take a different step than
\cite{Beisert:2004ry,Beisert:2003jj}, and follow ideas presented in
\cite{Zwiebel:2005er} -  we lift the result to finite $N$. First we
write spin chain states as a long trace
\begin{equation}
   \ket{X}\rightarrow \tr\left(\cdots (X^at^a)\cdots\right)~. \end{equation}
Then the generators \eqref{1loop+}-\eqref{1loop-} can be written
using matrix operators
\begin{subequations}
\begin{align}
    \mathfrak{T}^+_1
    =&+\frac1{\sqrt2}\sum_{k,q=0}^\infty
    \tr\cno{\psi_{(k)}\psi_{(q)}\check{\psi}_{(k+q+1)}}
\\
    \mathfrak{\bar T}^-_1
    =&+\frac1{\sqrt2}\frac1{N}\sum_{m,n=0}^\infty
    \left(\frac1{n+1}+\frac1{m+1}\right)
    \tr\cno{\psi_{(n+m+1)}\check{\psi}_{(n)}\check{\psi}_{(m)}}
\end{align}
\end{subequations}
$\check{\psi}_{(n)}$ is a subset of the ${\check\W}$ introduced
before (up to a different normalization in \eqref{parton-su11})
- it removes an operator $\psi_{(n)}$ and satisfies
$\left\{\check{\psi}_{(n)}^a,\psi_{(m)}^b\right\}:=\delta_{nm}\delta^{ab}$
where $a,b$ are adjoint indices of the gauge group. The colon's
stands for normal ordering i.e
\begin{equation*}
    \cno{\check\psi_{(n)}^a\psi_{(m)}^b} := \check\psi_{(n)}\psi_{(m)}-\delta_{nm}\delta^{ab} = -\psi_{(m)}\check\psi_{(n)}~.
\end{equation*}
Finally, the following $\SU(N)$ relation are useful
\begin{subequations}
\begin{align}
    \sum_{a}\tr At^a\tr B t^a =&\tr AB-\frac1{N}\tr A\tr B
\\
    \sum_{a}\tr At^aB t^a =&\tr A\tr B-\frac1{N}\tr AB~.
\end{align}
\end{subequations}
We now have the tools to calculate the $g^2$ correction to $\dD$ at
finite $N$ -
\begin{align}
    \dD_2 =& 2\left\{\T^+_1,\bar\T^-_1\right\}
    =\cr
    =
    &+\frac{1}{N}\sum_{m,n,k,q=0}^\infty\delta_{m+n=k+q}\left(\frac{\Theta(k-m)}{k-m}-\frac{\Theta(q-n)}{m+1}
    \right)\times\cr&\hspace{5em}\times
    \tr\cno{\left\{\psi_{(q)},\check\psi_{(n)}\right\}\left\{\psi_{(k)},\check\psi_{(m)}\right\}}+\cr
    &-\frac{1}{N}\sum_{m,n=0}^\infty\frac1{m+1}
    \tr\cno{\left\{\psi_{(m)},\check\psi_{(m)}\right\}\left\{\psi_{(n)},\check\psi_{(n)}\right\}}+\cr
    &+\sum_{m=0}^\infty 2h(m+1)\tr\cno{\psi_{(m+1)}\check\psi_{(m+1)}}~,
\end{align}
where the $\Theta$ are step functions
\begin{equation*}
    \Theta(x)=\begin{cases}
    0 & x\leq 0\\
    1 & x> 0
    \end{cases}
\end{equation*}
and $h(n)$ stands for the harmonic numbers
\begin{equation*}
    h(n):=\sum_{k=1}^{n}\frac1{k}~,\qquad h(0):=0~. \end{equation*}
This result agrees with the results of
\cite{Beisert:2004ry,Beisert:2003jj} if we note that
\begin{equation}
    \tr\cno{\psi_{(m)}\check\psi_{(m)}}
    =
    -\frac1{2N}
    \sum_{n=0}^\infty
    \tr\cno{\left\{\psi_{(m)},\check\psi_{(m)}\right\}\left\{\psi_{(n)},\check\psi_{(n)}\right\}}
    \label{newgg}
\end{equation}
since the difference between these operators is a gauge
transformation (see appendix \ref{app-gauge}), and we may use
\eqref{newgg} as an identity when acting on gauge invariant
operators. Using this property and some algebra, the
dilatation operator up to order $g^2$ is
\begin{align}
    \D =& \sum_{m=0}^\infty\left(\frac32+m\right)\tr\cno{\psi_{(m)}\check\psi_{(m)}}
    +g^2\Bigg[ \sum_{m=0}^\infty2h(m+1)\tr\cno{\psi_{(m)}\check\psi_{(m)}}+\cr
    &+\frac{1}{N}\sum_{m,k,q=0}^\infty\frac{(q+1)\Theta(k-m)}{(k-m)(k+q-m+1)}
    \tr\cno{\left\{\psi_{(k)},\check\psi_{(m)}\right\}\left\{\psi_{(q)},\check\psi_{(k+q-m)}\right\}}
    \Bigg]+\cr
    &+O(g^4)~.
\end{align}
The dilatation operator exhibit the form expected for a Fermi
liquid, the leading order in perturbation theory correct the mass
terms and add a 2-2 fermion interaction - in section \ref{sec-qp} we
will discuss the analogy in more details. For now we will calculate
the dilatation charge of the Fermi-sea state. The new ingredient in
the calculation is the 4-fermions coupling term. The annihilation
operators $\check\psi_{(m)}$ and $\check\psi_{(k+q-m)}$ puncture two
holes in the Fermi-sea, thus $m\leq K$ and $k+q-m\leq K$. Since the interaction
preserves momentum, at least one fermion creation operator must be
below (or equal to) the Fermi-level, but then Pauli's exclusion
principle forces it to fill one of the hole created, and we find
that the other creation operator must fill the other hole. To
summarize we have
\begin{align}
    \tr\cno{\left\{\psi_{(k)},\check\psi_{(m)}\right\}&\left\{\psi_{(q)},\check\psi_{(k+q-m)}\right\}}\ket{\Op^{(K)}_{\mathrm{1dim}}}
    \cr
    =&-\sum_{a,b=0}^{\dim G}\delta_{m=q}\tr\left[t^a,t^b\right]\left[t^b,t^a\right]\ket{\Op^{(K)}_{\mathrm{1dim}}}
    =\cr
    =&-2N(N^2-1)\delta_{q=m}\ket{\Op^{(K)}_{\mathrm{1dim}}}~, \end{align}
and for the dilatation operator we find
\begin{align}\label{g2dil}
    \D\ket{\Op^{(K)}_{\mathrm{1dim}}}=
    &(N^2-1)\sum_{m=0}^{K}\Bigg[\left(\frac32+m\right)
    +g^2\,2h(m+1)+\cr
    &-g^2\,\sum_{k=m+1}^{K}\frac{2(m+1)}{(k-m)(k+1)}
    +O(g^4)
    \Bigg]
    \ket{\Op^{(K)}_{\mathrm{1dim}}}=\cr
    =
    &(N^2-1)\frac{(K+3)(K+1)}{2}\left[1+\frac{4\,g^2}{K+3}+O(g^4)
    \right]\ket{\Op^{(K)}_{\mathrm{1dim}}}~. \end{align}

Note that the $g^2$ correction is suppressed by $1/K$ compared to the
leading order - this is the main point of this section. We will see
that this is also the case to order $g^4$. We would like to make two
comments that highlight the significance of this result:
\begin{itemize}
\item
We can compare this result to other operators which are familiar
from the spin-chain and integrability literature. For example,
consider another eigenstate of $\mathfrak{D}$, which is the analog of the bosonic
twist-2 operator (see \cite{Beisert:2004ry} for details)
\begin{align}
    \Op_{\mathrm{twist 2}}:=&\sum_{n=0}^{K}\frac{(-)^n(K)!}{n!(K-n)!}\tr\left(\psi_{(n)}\psi_{(K-n)}\right)
    \cr
    \D\ket{\Op_{\mathrm{twist 2}}} =&(K+3)\left[1+g^2\,\frac{2h(K+1)}{K+3}\right]\ket{\Op_{\mathrm{twist 2}}}
    \label{twista}
\end{align}
Using the asymptotics $h(K)\sim\log K$ for large $K$, we see that
the order $g^2$ ratio of the anomalous dimension over the classical
dimension scales, for large $K$, as $\sim   (\log K)/K$, whereas in our
case the $g^2$ correction is of order $1/K$.

\item Consider an operator in which we fill fermions states in a
sparse way up to some maximal level $K$. As long as the filling is
sparse such an operator can be written as a product of
\eqref{twista} operators. Since the $g^2$ anomalous dimension is a
two-to-two parton operator, the $\sim(\log K)/K$ will persist for such
sparse operators. Cancelation starts occurring as the Fermi shells
become more and more filled and Fermi statistics becomes the
dominant feature in the form of the operator.

\item We will see later that the cancelation of the logarithms
persists to order $g^4$, i.e., terms of the form $\sim N^2g^4 K (\log K)^2$
and $\sim N^2g^4 K\log K$ do not appear in the anomalous dimension. The
cancelation of the first set of terms, along with the cancelation of
$\sim N^2g^2 K\log K$ at this order, is important. Had these terms been
there, it would have suggested that the correction to the dimension
is of the form $\sim N^2K^{1+F(g^2)}$ which can become dominant at large
$g^2$. In our case, the cancelation of these terms suggests that the
correction to the dimension is of the form $\sim N^2KF(g^2)$ which is
parametrically smaller than the classical value of $N^2K^2$ even for
large $g^2$.

\end{itemize}

\subsection{The $g^4$ Hamiltonian}\label{sec-gs-2loop}

An algebraic solution for the order $g^4$ dilatation operator, based
on the $\psu(1,1|2)$ sector, was found by Zwiebel
\cite{Zwiebel:2005er}. This sector is generated by the partons
\begin{align}
    &\psi_{(n)} := \frac1{(n+1)!}\left(D_{1\dot1}\right)^n\Psi_{1\,4}
    &
    &\phi^2_{(n)} := \frac1{(n)!}\left(D_{1\dot1}\right)^n\Phi_{24}
    \cr
    &\bar\psi_{(n)} := \frac1{(n+1)!}\left(D_{1\dot1}\right)^n\bar\Psi^1_{\dot 1}
    &
    &\phi^3_{(n)} := \frac1{(n)!}\left(D_{1\dot1}\right)^n\Phi_{34}~.
\end{align}
Zwiebel's solution is reviewed in appendix \ref{app-2loop}. Here we quote the final result,
\begin{equation}\label{deltaD4}
    \dD_4=
    +2\left\{\bar\T^-_1,\left[\T^+_1,\left\{\T^-_1,\left[\bar\T^+_1,\mathfrak{h}\right]\right\}\right]\right\}
    +2\left\{\T^+_1,\left[\bar\T^-_1,\left\{\bar\T^+_1,\left[\T^-_1,\mathfrak{h}\right]\right\}\right]\right\}~,
\end{equation}
with
\begin{align}
    \mathfrak{h} :=& \sum_{n=0}^\infty
    +\frac12h(n+1)\tr\cno{\psi_{(n)}\check \psi_{(n)}}
    +\sum_{n=0}^\infty \frac12h(n+1)\tr\cno{\bar\psi_{(n)}\check{\bar\psi}_{(n)}}
    +\cr
    &+\sum_{n=0}^\infty \frac12h(n)\sum_{i=2}^3\tr\cno{\phi^i_{(n)}\check \phi^i_{(n)}}~,
\end{align}
and $\T_1^\pm$ and $\bar\T_1^\pm$ are the leading order ($g^1$)
supercharges of the $\psu(1|1)^2$ algebra, which commutes with the
$\psu(1,1|2)$ algebra (which can be found by a generalization of the calculation for $\su(1,1)$ fermionic sector)
\begin{subequations}\label{supercharges}
\begin{align}
    \T^+_1
    =&\frac1{\sqrt2}\sum_{k,q=0}^\infty
    \biggl(\tr\cno{\psi_{(k)}\psi_{(q)}\check{\psi}_{(k+q+1)}}
    +\sum_{i=2}^3\tr\cno{\left[\psi_{(k)},\phi^i_{(q)}\right]\check{\phi}^i_{(k+q+1)}}
    +\cr&\hspace{4em}
    +\frac{q+1}{(k+q+2)}\tr\cno{\left\{\psi_{(k)},\bar\psi_{(q)}\right\}\check{\bar\psi}_{(k+q+1)}}
    +\cr&\hspace{4em}
    -\frac1{k+q+1}\tr\cno{\left[\phi^2_{(k)},\phi^3_{(q)}\right]\check{\bar\psi}_{(k+q)}}
    \biggr)
\\
    \bar\T^+_1
    =&\frac1{\sqrt2}\sum_{k,q=0}^\infty\biggl(
    \tr\cno{\bar\psi_{(k)}\bar\psi_{(q)}\check{\bar\psi}_{(k+q+1)}}
    +\sum_{i=2}^3\tr\cno{\left[\bar\psi_{(k)},\phi^i_{(q)}\right]\check{\phi}^i_{(k+q+1)}}
    +\cr&\hspace{4em}
    +\frac{q+1}{(k+q+2)}\tr\cno{\left\{\bar\psi_{(k)},\psi_{(q)}\right\}\check{\psi}_{(k+q+1)}}
    +\cr&\hspace{4em}
    +\frac1{k+q+1}\tr\cno{\left[\phi^2_{(k)},\phi^3_{(q)}\right]\check{\psi}_{(k+q)}}
    \biggr)
\\
    \T^-_1
    =&\frac1{\sqrt2N}\sum_{m,n=0}^\infty\biggl(
    \frac{n+m+2}{(n+1)(m+1)}\tr\cno{\bar\psi_{(n+m+1)}\check{\bar\psi}_{(n)}\check{\bar\psi}_{(m)}}
    +\cr&\hspace{4em}
    -\tr\cno{\psi_{(n+m)}\left[\check{\phi}^2_{(n)},\check{\phi}^3_{(m)}\right]}
    +\frac1{n+1}
    \tr\cno{\psi_{(n+m+1)}\left\{\check{\bar\psi}_{(n)},\check{\psi}_{(m)}\right\}}
    +\cr&\hspace{4em}
    +\frac1{n+1}\sum_{i=2}^3
    \tr\cno{\phi^i_{(n+m+1)}\left[\check{\bar\psi}_{(n)},\check{\phi}^i_{(m)}\right]}
    \biggr)
\\
    \bar\T^-_1
    =&\frac1{\sqrt2N}\sum_{m,n=0}^\infty\biggl(
    \frac{n+m+2}{(n+1)(m+1)}
    \tr\cno{\psi_{(n+m+1)}\check{\psi}_{(n)}\check{\psi}_{(m)}}
    +\cr&\hspace{4em}
    +\tr\cno{\bar\psi_{(n+m)}\left[\check{\phi}^2_{(n)},\check{\phi}^3_{(m)}\right]}
    +\frac1{n+1}
    \tr\cno{\bar \psi_{(n+m+1)}\left\{\check{\psi}_{(n)},\check{\bar \psi}_{(m)}\right\}}
    +\cr&\hspace{4em}
    +\frac1{n+1}\sum_{i=2}^3
    \tr\cno{\phi^i_{(n+m+1)}\left[\check{\psi}_{(n)},\check{\phi}^i_{(m)}\right]}
    \biggr)~.
\end{align}
\end{subequations}

There are two caveats in Zwiebel's solutions (already discussed in
\cite{Zwiebel:2005er}): (1) It is not clear that the solution is
unique. In general there could exist a "homogenous" set of
deformations of the generators (beside the coupling redefinition and
similarity transformation) which do not effect the commutation
relations. No such "homogeneous solutions" are known though. (2) The
lift to the non-planarity is not proven to be unique. Some checks at large N
(by comparison to Feynman diagrams) of the solution for small
spin chain operators are provided in
\cite{Zwiebel:2005er}. The appearance of a Yangian symmetry
\cite{Zwiebel:2006cb} also supports the solution, at least in the
spin chain limit. For more recent discussion about the validity of
the solution beyond the planar limit see \cite{Zwiebel:2008gr}.

\begin{figure}[ht]
  \begin{center}
    \mbox{
      \subfigure[The ${\bar\T}^+_1$ vertices\label{fig1}]{\scalebox{0.45}{\includegraphics[width=1\textwidth]{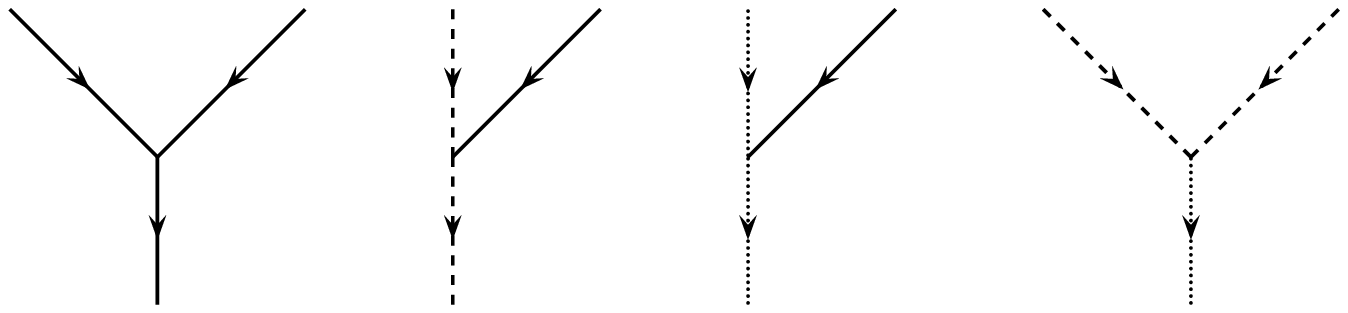}}}
      \qquad\quad
      \subfigure[The ${\bar\T}^-_1$ vertices\label{fig2}]{\scalebox{0.45}{\includegraphics[width=1\textwidth]{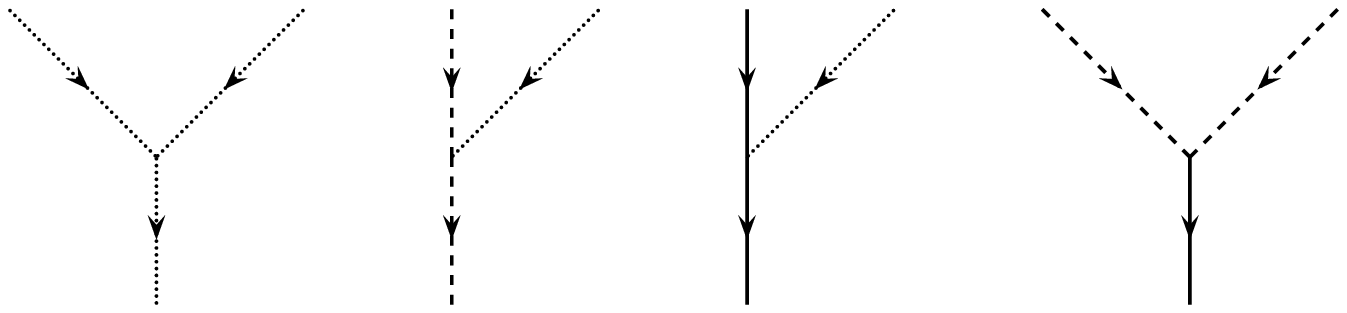}}}
      }
    \caption{The continuous, dashed, doted lines stands for $\psi_{(n)}$,
$\phi^i_{(n)}$ and $\bar\psi_{(n)}$ respectively. The arrows
indicate the flow of momentum. For the conjugated diagrams
${\bar\T}^-_1$ and ${\T}^+_1$ the vertices structure do not change,
only the arrows changes sign. Note that these are not Feynman
diagrams, but rather they only describe the how the various terms in
these operators act. \label{charges} }
  \end{center}
\end{figure}

The supercharges $\T^\pm_1$ and $\bar\T^\pm_1$ can be described
diagrammatically as a set of vertices (see figure \ref{charges}) -
the $\T^\pm_1$ split a single parton into a pair of partons and the
$\bar\T^\pm_1$ combine a pair of parton into a single parton. For
our purpose, it is enough to focus on the fermionic $\su(1,1)$ sector,
and examine how each operator acts on the sector.

\begin{figure}[ht]
  \begin{center}
    \mbox{
      \subfigure{\scalebox{0.30}{\includegraphics[height=1.2\textwidth]{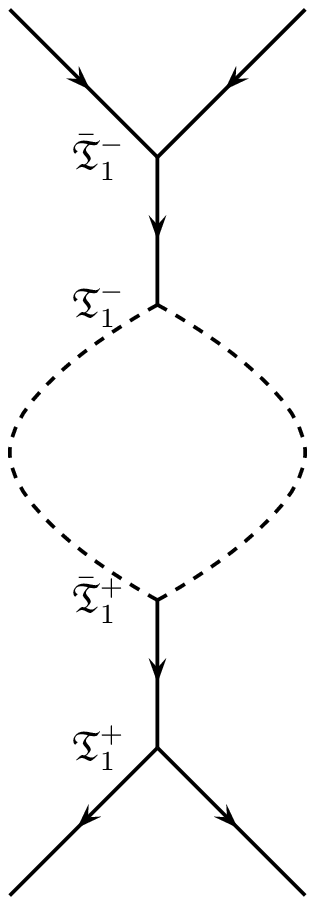}}}
      \qquad
      \subfigure{\scalebox{0.30}{\includegraphics[height=1.2\textwidth]{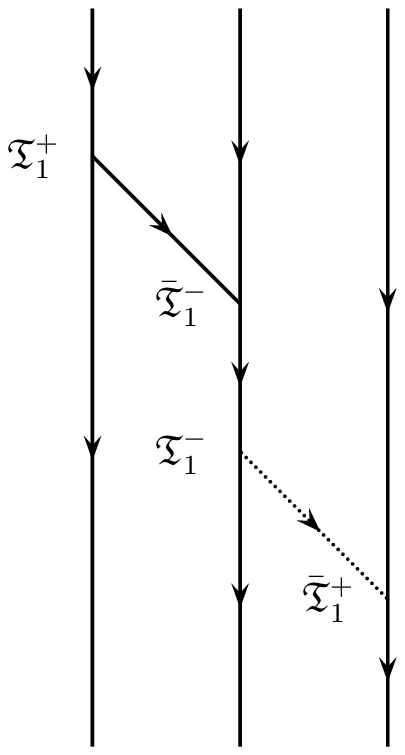}}}
      \qquad
      \subfigure{\scalebox{0.30}{\includegraphics[height=1.2\textwidth]{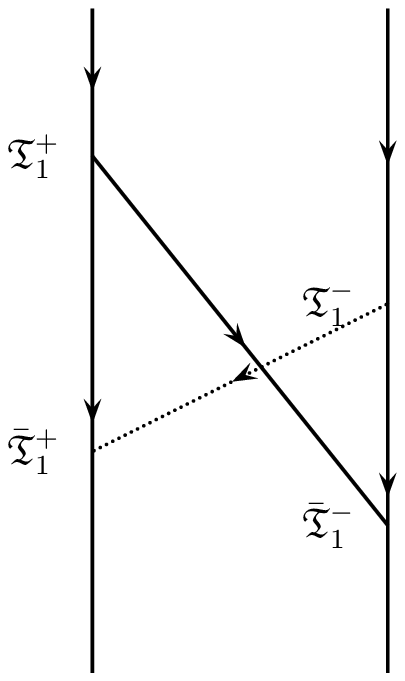}}}
      }
        \caption{Three typical diagrams for $\dD_4$, which keep track of how
the partons change when acting on them with the sequence of
operators in (3.26). The vertices are ordered in the vertical axis
by the order the operators acts on the state. Since
$\T^-~,~\bar\T^+_1$ appear in the inner commutators only, the
corresponding vertices are inserted sequently, as a result they form
a either a 1-1 and 2-2 fermionic operator, any none fermionic line
emerging from the inner commutator will continue to the outer legs
(irrelevant for our purpose). \label{sample}}
  \end{center}
\end{figure}

First consider the contribution of the operators
$\T^-_1~,~\bar\T^+_1$ in \eqref{deltaD4}. Since $\mathfrak{h}$ is a
1-1 parton operator, which cannot change the parton, the
combinations
$\left\{\T^-_1,\left[\bar\T^+_1,\mathfrak{h}\right]\right\}$ and
$\left\{\bar\T^+_1,\left[\T^-_1,\mathfrak{h}\right]\right\}$ have
the same structure as $\dD_2$ (the difference is in coefficients
only). Since $\dD_2$ acts within the fermionic $\su(1,1)$ sector,
these combinations acts within the fermionic $\su(1,1)$ sector too.
The operators $\T^+_1~,~\bar\T^-_1$ and $\dD_2$ also act within the
fermionic $\su(1,1)$ sector. Hence, after we calculate the inner
commutators (involving $\T^-_1~,~\bar\T^+_1$ and $\mathfrak{h}$) in
the full $\psu(1,1|2)$ sector we can continue the calculation in the
smaller $\su(1,1)$ sector. Some sample diagrams, contributing to
$\dD_4$, which illustrate the argument are sketched in figure
\ref{sample} (these are not Feynman diagrams. These diagrams only
keep track of the partons and of the ordering of how the operators
act on them).

We first evaluate $\dD_4$ on the $\su(1,1)$ sector, and then use the
resulting expression to evaluate it on
$\ket{\Op_{\mathrm{1dim}}^{(K)}}$. First we take care of the inner
commutators
\begin{subequations}
\begin{align}
    2\left\{\mathfrak{\bar T}^+_1,\left[\mathfrak{T}^-_1,\mathfrak{h}\right]\right\}
    =&\mathfrak{U}-\mathfrak{G}
    \\
    2\left\{\mathfrak{T}^-_1,\left[\mathfrak{\bar T}^+_1,\mathfrak{h}\right]\right\}
    =&\mathfrak{G}-\mathfrak{V}~,
\end{align}
\end{subequations}
with
\begin{subequations}
\begin{align}
    \mathfrak{U}
    =&\frac1{2N}\sum_{\substack{m,n,\\k,q=0}}^\infty\delta_{m+n=k+q}\mathcal{A}_{(m,n-q-1)}\mathcal{B}_{[n-q-1,n]}
    \tr\cno{\left\{\psi_{(k)},\check\psi_{(m)}\right\}\left\{\psi_{(q)},\check\psi_{(n)}\right\}}
    \\
    \mathfrak{V}
    =&\frac1{2N}\sum_{\substack{m,n,\\k,q=0}}^\infty\delta_{m+n=k+q}\mathcal{A}_{(q,n-q-1)}\mathcal{B}_{[n-q-1,n]}
    \tr\cno{\left\{\psi_{(k)},\check\psi_{(m)}\right\}\left\{\psi_{(q)},\check\psi_{(n)}\right\}}
    \\
    \mathfrak{G}
    =&-\frac12\sum_{\substack{m=0}}^\infty\mathcal{D}_{(m)}\tr\cno{\psi_{(m)}\check\psi_{(m)}}~,
\end{align}
\end{subequations}
where the constants are
\begin{subequations}
\begin{align}
    \mathcal{A}_{(a,b)}=&\frac{h(a+1)+h(b+1)-h(a+b+2)}{2}\Theta(a+1)\Theta(b+1)
    \\
    \mathcal{B}_{[a,b]}=&\frac{b-a}{(a+1)(b+1)}\Theta(a+1)\Theta(b+1)
    \\
    \mathcal{C}_{(a,b)}=&\frac{a+b+2}{(a+1)(b+1)}\Theta(a+1)\Theta(b+1)
    \\
    \mathcal{D}_{(a)}= &-\frac{h(a+1)}{(a+1)}
    +\Theta(a)\sum_{b=0}^{a-1}\frac{h(b+1)+h(a-b)-h(a+1)}{a-b}
\end{align}
\end{subequations}
Next we calculate the $g^3$ terms
\begin{subequations}
\begin{align}
    \left[\mathfrak{T}^+_1,\mathfrak{G}\right]=
    &\frac1{4\sqrt2}\sum_{k,q=0}^\infty\,\xi_{(k,q)}\,
    \tr\cno{\left\{\psi_{(k)},\psi_{(q)}\right\}\check\psi_{(k+q+1)}}
\\
    \left[\mathfrak{\bar T}^-_1,\mathfrak{G}\right]=
    &-\frac1{4\sqrt2\,N}\sum_{k,q=0}^\infty\,\mathcal{C}_{(k,q)}\xi_{(k,q)}
    \,
    \tr\cno{\psi_{(k+q+1)}\left\{\check\psi_{(k)},\check\psi_{(q)}\right\}}
\\
    \left[\mathfrak{T}^+_1,\mathfrak{V}\right]=
    &\frac1{4\sqrt2\,N}\sum_{\substack{m,n,\\q,k=0}}^\infty\,
    \nu_{[(q;n),(k;m)]}\times
    \cr
    &\hspace{4em}\times
     \tr\cno{\left[\left\{\psi_{(q)},\check\psi_{(n)}\right\},\left\{\psi_{(k)},\check\psi_{(m)}\right\}\right]\psi_{(m+n-k-q-1)}}
    +\cr
    &+\frac1{4\sqrt2}\sum_{\substack{k,q=0}}^\infty\bar\nu_{(k,q)}
    \tr\cno{\left\{\psi_{(q)},\psi_{(k)}\right\}\check\psi_{(k+q+1)}}
\\
    \left[\mathfrak{\bar T}^-_1,\mathfrak{U}\right]=
    &\frac1{4\sqrt2\,N^2}\sum_{\substack{m,n,\\k,q=0}}^\infty
    \mu_{[(q;n),(k;m)]}\times
    \cr
    &\hspace{4em}\times
    \tr\cno{\left[\left\{\psi_{(q)},\check\psi_{(n)}\right\},\left\{\psi_{(k)},\check\psi_{(m)}\right\}\right]\check\psi_{(k+q-m-n-1)}}
    +\cr
    &-\frac1{4\sqrt2\,N}\sum_{\substack{k,q=0}}^\infty\bar\mu_{(k,q)}\,
    \tr\cno{\psi_{(k+q+1)}\left\{\check\psi_{(k)},\check\psi_{(q)}\right\}}
\end{align}
\end{subequations}
In the above calculation we use the following identity proved in appendix \ref{app-gauge}
\begin{multline}\label{gid2}
    \tr\cno{\left\{\psi_{(k)},\psi_{(q)}\right\}\check\psi_{(k+q+1)}}
    \widehat{=}\\\widehat{=}\,
    -\frac1{2N}\sum_{m=0}^\infty\tr\cno{\left\{\psi_{(m)},\check\psi_{(m)}\right\}
    \left[\left\{\psi_{(q)},\psi_{(k)}\right\},\check\psi_{(k+q+1)}\right]}
\end{multline}
and define
\begin{subequations}
\begin{align}
    \mu_{[(q;n)(k;m)]}:=\,&
    \Theta(k+q-m-n)\times\cr
    \times\Big[&
    A_{(m,k-m-1)}B_{[k+q-m-n-1,k-m-1]}C_{[n-q-1,n]}
    +\cr&
    -A_{(k+q-n,n-q-1)}B_{[k+q-m-n-1,k-m-1]}C_{[n-q-1,n]}
    +\cr&
    -A_{(n,q-n-1)}B_{[k+q-m-n-1,q-n-1,]}C_{[m-k-1,m]}
    +\cr&
    -A_{(k+q-m,m-k-1)}B_{[k+q-m-n-1,q-n-1,]}C_{[m-k-1,m]}\Big]
\\
    \nu_{[(q;n)(k;m)]}:=\,&
    \Theta(m+n-k-q)\times\cr
    \times\Big(&
    A_{(m+n-k,k-m-1)}B_{[k-m-1,n]}
    -A_{(m+n-k-q-1,k-m-1)}B_{[k-m-1,n-q-1]}+\cr
    &
    -A_{(m+n-q,q-n-1)}B_{[q-n-1,m]}
    +A_{(m+n-k-q-1,q-n-1)}B_{[q-n-1,m-k-1]}+\cr
    &
    -\Theta(k-m+1)A_{(q,n-q-1)}B_{[n-q-1,n]}+\cr
    &
    +\Theta(q-n+1)A_{(k,m-k-1)}B_{[m-k-1,m]}
    \Big)
\\
    \bar\nu_{(k,q)}:=\,&
    \sum_{n=0}^{k+q}\left(\mathcal{A}_{(q,n-q-1)}\mathcal{B}_{[n-q-1,n]}+\mathcal{A}_{(k,n-k-1)}\mathcal{B}_{[n-k-1,n]}\right)
\\
    \bar\mu_{(k,q)}:=\,&
    \sum_{n=0}^{k+q}\left(\mathcal{A}_{(q,k-n-1)}\mathcal{B}_{[k-n-1,k]}+\mathcal{A}_{(k,q-n-1)}\mathcal{B}_{[q-n-1,q]}\right)\mathcal{C}_{(k+q-n,n)}
\\
    \Gamma(k,q):=&\mathcal{D}_{(k)}+\mathcal{D}_{(q)}-\mathcal{D}_{(k+q+1)}
\end{align}
\end{subequations}

The notations in the above are such that the brackets $(a,b)$ indicates a
symmetry replacing the momenta between them, the square brackets
$[a,b]$ indicates antisymmetry, and brackets with a semicolon
$(a;b)$ has no symmetry. The last stage is calculating $\dD_4$, for
which we can use
\begin{align}
    \dD_4
    =
    \left\{\mathfrak{\bar T}^-_1,\left[\mathfrak{T}^+_1,\mathfrak{G}\right]\right\}
    -\left\{\mathfrak{\bar T}^-_1,\left[\mathfrak{T}^+_1,\mathfrak{V}\right]\right\}
    -\left\{\mathfrak{T}^+_1,\left[\mathfrak{\bar T}^-_1,\mathfrak{G}\right]\right\}
    +\left\{\mathfrak{T}^+_1,\left[\mathfrak{\bar
    T}^-_1,\mathfrak{U}\right]\right\}~. \end{align}
Summing all the contributions we find
\begin{align}\label{deltaD4-operator}
    \dD_4
    =
    &+\frac1{4N^2}\sum_{\substack{m,n,k,q,\\r,s,t,l=0}}^\infty
    \mu_{[(q;n),(k;m)]}\delta_{l=k+q-m-n-1}\delta_{t=r+s+1}\times\cr
    &\quad\times\Bigg[
    -\delta_{s=n}
    \tr\cno{\left[\left\{\psi_{(k)},\check\psi_{(m)}\right\},\check\psi_{(l)}\right]\left[\left\{\psi_{(r)},\check\psi_{(t)}\right\},\psi_{(q)}\right]}
    +\cr&\hspace{3em}
    +\delta_{t=q}    \tr\cno{\left[\left\{\psi_{(k)},\check\psi_{(m)}\right\},\check\psi_{(l)}\right]\left[\left\{\psi_{(r)},\check\psi_{(n)}\right\},\psi_{s}\right]}
    +\cr&\hspace{3em}
    -\delta_{r=l}
    \tr\cno{\left[\left\{\psi_{(k)},\check\psi_{(m)}\right\},\check\psi_{(t)}\right]\left[\left\{\psi_{(q)},\check\psi_{(n)}\right\},\psi_{(s)}\right]}
    +\cr&\hspace{3em}
    -\delta_{s=n}\delta_{r=l}N\tr\cno{\left\{\psi_{(k)},\check\psi_{(m)}\right\}\left\{\psi_{(q)},\check\psi_{(t)}\right\}}
    +\cr&\hspace{3em}
    +\delta_{s=n}\delta_{r=m}\bigg(N\tr\cno{\psi_{(q)}\check\psi_{(t)}\psi_{(k)}\check\psi_{(l)}}
    +2\tr\cno{ \psi_{(q)}\check\psi_{(t)}}\tr\cno{\psi_{(k)}\check\psi_{(l)}}
    +\cr&\hspace{3em}
    +\tr\cno{\psi_{(q)}\psi_{(k)}}\tr\cno{\check\psi_{(l)}\check\psi_{(t)}}
    \bigg)
    \Bigg]
    +\cr
    &+\frac1{4N^2}\sum_{\substack{m,n,k,q,\\r,s,l,t=0}}^\infty
    C_{(r,s)}\nu_{[(q;n),(k;m)]}\delta_{l=m+n-k-q-1}\delta_{t=r+s+1}\times\cr
    &\quad\times\Bigg[
    -\delta_{s=q}
    \tr\cno{\left[\left\{\psi_{(t)},\check\psi_{(r)}\right\},\check\psi_{(n)}\right]\left[\left\{\psi_{(k)},\check\psi_{(m)}\right\},\psi_{(l)}\right]}
    +\cr&\hspace{3em}
    +\delta_{t=n}
    \tr\cno{\left[\left\{\psi_{(q)},\check\psi_{(r)}\right\},\check\psi_{(s)}\right]\left[\left\{\psi_{(k)},\check\psi_{(m)}\right\},\psi_{(l)}\right]}
    +\cr&\hspace{3em}
    -\delta_{r=l}
    \tr\cno{\left[\left\{\psi_{(q)},\check\psi_{(n)}\right\},\check\psi_{(s)}\right]\left[\left\{\psi_{(k)},\check\psi_{(m)}\right\},\psi_{(t)}\right]}
    +\cr&\hspace{3em}
    -\delta_{s=q}\delta_{r=l}N
    \tr\cno{\left\{\psi_{(t)},\check\psi_{(n)}\right\}\left\{\psi_{(k)},\check\psi_{(m)}\right\}}
    +\cr&\hspace{3em}
    +\delta_{s=q}\delta_{r=k}\bigg(
    +N\tr\cno{\psi_{(t)}\check\psi_{(n)}\psi_{(l)}\check\psi_{(m)}}
    +2\tr\cno{\psi_{(t)}\check\psi_{(n)}}\tr\cno{\psi_{(l)}\check\psi_{(m)}}
    +\cr&\hspace{3em}
    +\tr\cno{\psi_{(t)}\psi_{(l)}}\tr\cno{\check\psi_{(m)}\check\psi_{(n)}}
    \bigg)
    \Bigg]
    +\cr
    &+\frac1{2N}\sum_{r,s,k,q=0}^\infty
    \left(C_{(k,q)}\xi_{(r,s)}+C_{(k,q)}\xi_{(k,q)}
    -\bar\mu_{(k,q)}
    -C_{(k,q)}\bar\nu_{(r,s)}
    \right)\times\cr
    &\quad\times
    \bigg(
    -\frac12\delta_{k+q=r+s}\tr\cno{\left\{\psi_{(s)},\check\psi_{(k)}\right\}\left\{\psi_{(r)},\check\psi_{(q)}\right\}}
    +\cr&\qquad\quad
    +\delta_{r=k}\tr\cno{\left\{\psi_{(s)},\check\psi_{(k+s+1)}\right\}\left\{\psi_{(k+q+1)},\check\psi_{(q)}\right\}}
    +\cr&\qquad\quad
    +\delta_{s=k}\delta_{r=q}N\tr\cno{\psi_{(k+q+1)}\check\psi_{(k+q+1)}}\bigg)
\end{align}

We are set to apply the dilatation operator to the Fermi-sea. The details are
given in appendix \ref{app-g4-Fermi}, and here we only quote the
result in the large $K$ and large $N$ limit
\begin{equation}
    \D\ket{\Op^{(K)}_{\mathrm{1dim}}}
    =
    N^2\frac{K^2}{2}\left[1+\frac{4\,g^2}{K}-\frac{4g^4}{K} + O(K^{-2})+O(g^{6})\right]\ket{\Op^{(K)}_{\mathrm{1dim}}}~.
    \label{findim4}
\end{equation}
This is the main result of the paper, as far as the explicit
computation of the dimension is concerned.

Before proceeding to analyze the result it is important to emphasize
that the expression in \eqref{findim4} contains contributions from
non-planar diagrams which are not suppressed by power of N compared
to the planar ones. This, however, is not in contradiction to the
standard lore, since the operator has a strong N dependence as it
has order $N^2K$ partons (stronger in fact than determinants/giant
gravitons \cite{McGreevy:2000cw,Balasubramanian:2001nh}).

We can also repeat the analysis that we had before, examining which terms cancel, and compare to the bosonic twist
operator. At order $g^4$ there are $K(\log K)^2$ and $K\log K$
coefficients which vanish (the significance of this was discussed
after equation \eqref{g2dil}).
Compared to the bosonic twist 2 operator, the latter has a leading $\log$ behavior
\begin{multline}
    \D\ket{\tr\left(\Phi_{14}(D_{1\dot 1})^s \Phi_{14}\right)}=\\
    =s\left[1+ f(g)\frac{\log s}{s}+h(g)\frac{\log^2 s}{s} + O(s^0)\right]\ket{\tr\left(\Phi_{14}(D_{1\dot 1})^s M_{14}\right)}
\end{multline}
with $f(g)=4g^2-\frac{2}{3}\pi^2g^4+\cdots$ (replacing the scalar in
the above with a fermion will not make a considerable change of the
result).
As before, the situation for the Fermi-sea is better than the
expected expansion in $g^{2n}(\log K)^n$.
Rather, the persistence of the cancelation to order $g^4$ suggests
that the behavior at large $K$ is
\begin{equation}\label{fgsq}
    \D\ket{\Op^{(K)}_{\mathrm{1dim}}}
    =N^2\frac{K^2}{2}\left[1+\frac{F(g^2)}{K}+ O(K^{-2})\right]\ket{\Op^{(K)}_{\mathrm{1dim}}}~.
\end{equation}

The function $F(g)$ is analogous to the famous cusp anomalous
dimension calculated to all order by Beisert, Eden and Staudacher
\cite{Beisert:2006ez,Eden:2006rx} for the twist 2 operators. \footnote{See also \cite{Freyhult:2007pz} for a generalization to twist-j operators.} From the work of
Frolov and Tseytlin \cite{Frolov:2002av}, we know that the anomalous
dimension of a twist 2 operator remains finite at strong coupling
($g\rightarrow\infty$), as it is dual to a folded spinning string
\cite{Gubser:2002tv}. As the situation for the Fermi-sea operator
seems somewhat better - in the sense that the corrections are
relatively smaller - one can hope that this class of operators
continues into to strong coupling, where they will be 1/16 BPS
states, up to small corrections.

\section{Quasi particles}\label{sec-qp}

Next, we study the quasi-particles - the small excitations above the
Fermi surface - to order $g^2$. We will not compute all excitations
of the Fermi surface by all operators in the CFT, but rather only
excitations in the fermionic $\su(1,1)$ sector. The starting point
is the free-energy restricted to this sector
\begin{multline}\label{fenrgy}
    \F = \dD = g^2\sum_{k=0}^\infty\Bigg[2h(k+1)\sum_a\psi^a_{(k)}\check\psi^a_{(k)}
    +\\+
    \frac{1}{N}\sum_{p=0}^\infty\sum_{q=-\infty}^\infty
    \sum_{a,b,c,d,e}\frac{(p+1)\Theta(k-p+q)}{(k-p+q)(k+q+1)}f_{ab}^ef_{cd}^e
    \psi_{(k)}^a\psi_{(p)}^c\check\psi_{(p-q)}^b\check\psi_{(k+q)}^d
    \Bigg]~, \end{multline}
where we made a change of the summation variables to a form more
familiar in condensed matter, the traces and commutator are written
using the $\SU(N)$ structure constants. Define the fermions density
\begin{equation}
    n^a(k):=\ave{\psi_{(k)}^a\check\psi_{(k)}^a}~. \end{equation}
We study the quasi-particle using a Hartree-Fock approximation \footnote{See \cite{Nozires:1966aa} for the application of a similar approximation in the microscopic study of normal Fermi liquids in condensed matter physics.}
\begin{equation}
    E = \ave{\F}\spc
    \ave{\psi_{(k)}^a\check\psi_{(p)}^b}=\delta^{ab}\delta_{kp}n^a(k)~. \end{equation}
Applying the Hartree-Fock approximation to \eqref{fenrgy}, we find
that the energy in term of the density is
\begin{multline}\label{enrgy}
    E = g^2\sum_{k=0}^\infty\Bigg[2h(k+1)\sum_an^a(k)
    +\\+
    \frac{1}{N}\sum_{p=0}^{\infty}
    \sum_{a,b,e}\frac{(p+1)\Theta(k-p)}{(k-p)(k+1)}f_{ab}^ef_{ba}^en^b(p)n^a(k)
    \Bigg]~. \end{multline}
The quasi particle energy ($\dE$) and distribution ($\dn$) are defined by removing the equilibrium energy and distribution
\begin{equation}
    \dE = E-\bar E
    \spc
    \dn^a(q):=\ave{\psi_{(q)}^a\check\psi_{(q)}^a}-\bar n^{a}(q)~,
\end{equation}
where the equilibrium state for the 1-dim Fermi-sea is
\begin{equation}
    \bar n^{a}(k)=1-\Theta(k-K)~. \end{equation}

Applying the above to \eqref{enrgy}, we find the quasi particle
energy in term of their distribution function
\begin{align}
    \dE = g^2\Bigg[&
    \sum_{k=0}^\infty\sum_{a}\bigg(
    2h(k+1)-2\sum_{p=0}^{K}\frac{(p+1)\Theta(k-p)}{(k-p)(k+1)}
    +\cr&
    -2\sum_{p=0}^{K}\frac{(k+1)\Theta(p-k)}{(p-k)(p+1)}\bigg)\dn^a(k)
    +\cr&+
    \frac{1}{N}\sum_{k,p=0}^\infty
    \sum_{a,b,e}\frac{(p+1)\Theta(k-p)}{(k-p)(k+1)}f_{ab}^ef_{ba}^e\dn^b(p)\dn^a(k)
    \Bigg]
\end{align}
(where negative $\delta n$ for momenta below the Fermi surface are
the quasi-holes).

 The
quasi-particle energy should be measured relatively to the
Fermi-level. We denote the momentum above the Fermi-surface as
\begin{equation*}
    k:= K+e~. \end{equation*}
The single quasi-particle energy (for a particle above the Fermi-sea $e>0)$ is
\begin{multline}\label{qp-single}
    \varepsilon_{q.p}(e) = 2g^2\left[\,h(K+e+1)-\sum_{p=0}^{K}\frac{(p+1}{(K+e-p)(K+e+1)}\right]
    =\\
    =\frac{g^2}{2}\left(\frac{K+2}{K+e+1}+h(e-1)\right)
    \approx \frac{g^2}{2}\left[1+h(e-1)-\frac{e-1}{K}O(K^{-2})\right]~. \end{multline}
For completeness we also calculate the single quasi-hole energy \footnote{For a quasi-hole the energy is $\varepsilon_{q.h}(e):=-\varepsilon_{q.p}(-e)$ with $e>0$.} (i.e a particle below the Fermi-sea) :
\begin{multline}
    \varepsilon_{q.h}(e) = -2g^2\Bigg[\,h(K-e+1)-\sum_{p=0}^{K-e-1}
    \frac{(p+1)}{(K-e-p)(K-e+1)}+\\-\sum_{p=K-e+1}^{K}\frac{(K-e+1)}{(p-K+e)(p+1)}\Bigg]
    \approx \frac{g^2}{2}\left[1+h(e)-\frac{2+e}{K}+O(K^{-2})\right]~. \end{multline}

Note that contributions to these energies come from both the
quadratic piece and the interaction piece in \eqref{fenrgy}. This is
crucial as the leading $\log K$ contribution to the energy of the
quasi-particle cancels between these two contributions terms. To see
this, we note that the quadratic term in \eqref{fenrgy} by itself
gives a contribution which is $\varepsilon_{q.p}(e)=2g^2
h(K+e+1)\sim g^2 \log K$.

The fact that we found a much smaller energy suggests that the same
cancelation mechanism, found in section \ref{sec-gs} for the ground
state energy, is also present for the excited states. One might
therefore hope that the quasi-particles could eventually be
understood on the gravity side as well. A caveat to the analysis
that we carried out here is that we computed only the self energy
piece - i.e. the diagonal element in the Hamiltonian in this state -
and did not solve the entire mixing problem. The standard lore in
Fermi liquids, however, is that mixing is possible but that the
lifetime of the quasi-particles increases as the energy decreases.
Of course, it could also be that the situation changes at higher
orders of $g^2$.

\section{A short discussion about 2-dim Fermi-sea}\label{sec-2d}

In \cite{Berkooz:2006wc} we were interested in semi-short states,
invariant under a single supersymmetry charge (1/16-BPS). At the
leading $g$ behavior the supercharge changes a covariant derivative
to fermion, this can be written as
\begin{equation*}
    \left\{\bar Q_{24},\psi_{(n)}\right\}
    \sim \sum_{m=0}^{n-1}\left\{\psi_{(m)},\psi_{(n-1+m)}\right\}
\end{equation*}
When considering only the action of the supercharge we constructed a
2-dim Fermi-sea operator by relaxing the condition $j_L=j_R-q_2$~,
for example one can have $j_L=0$ which is the one that we will
focus on. The construction is similar to the 1-dim case. Relaxing
the condition on the left-handed angular moment (undotted indices)
the relevant fermionic partons are
\begin{align}\label{parton-su12}
    \psi_{(n|0)} :=&
    \frac1{(n+1)!}(\ab_1^\dag)^{n+1}(\bb_{\dot1}^\dag)^{n}\cb_4^\dag\ket{0}
\cr
    \psi_{(n|m)} :=&
    \frac1{(n+1-m)!(m)!}(\ab_1^\dag)^{n+1-m}(\ab_2^\dag)^{m}(\bb_{\dot1}^\dag)^{n}\cb_4^\dag\ket{0}
    =(\J_L^-)^m\psi_{(n|0)}~,
\end{align}
where $\J_L^-$ is the $\SU(2)_L$ operator which lower the $j_L^3$
charge. It is important to notice that all left handed angular
momentum indices are symmetrized for a single fermion\footnote{An anti-symmetrization should be replaced, using the equation of motion, to a commutator of partons:~
$\epsilon^{\alpha\beta}\left[D_{\alpha\dot1},D_{\alpha\dot1}\right]=2\bar
F_{\dot1\dot1}$ and
$\epsilon^{\alpha\beta}D_{\alpha\dot1}\Psi_{\beta}=\left[\bar\Psi_{\dot1}^i,\Phi_{i4}\right]$. We prevent this by forcing the discussed symmetrization}.
The level of the fermion is $n$ and it's angular momentum is
\begin{equation*}
    \left(j_L,j_L^3~;~j_R,j_R^3\right)
    =\left(\frac{n+1}{2},\frac{n+1}2-m~ ;~ \frac{n}{2},\frac{n}2\right)~. \end{equation*}
Notice that the components of the momentum vector obey
\begin{equation*}
    \left(\frac{n+1}2\right)^2-\left(\frac{n+1}2-m\right)^2\geq 0~.
\end{equation*}
Thus the momentum is confined to a 'forward light-cone', which is
where the 'relativistic' nature of the Fermi-sea comes from. The
$j_L=0$ state can be built by multiplying all fermions with all
values of $j_L^3$. This is the same as multiplying all fermions with
as few derivative as possible - notice that now at each level there
is a degeneracy due to different momentum vectors. Hence the
$j_L=0$ operator is
\begin{equation}\label{Fermi-sea-2dim}
    \Op^{(K)}_{\mathrm{2dim}}
    :=\prod_{n=0}^{K}\prod_{m=0}^{n+1}\Jdet\left(\psi_{(I|m)}\right)
    =\prod_{n=0}^{K}\prod_{m=0}^{n+1}\Jdet\left(\frac1{m!}(\J_L^-)^m\psi_{(I|0)}\right)~. \end{equation}
The energy \footnote{A straightforward computation gives
$q_2\sim N^2 K^2/2$ and $j_R\sim N^2K^3/6$, corresponding to the bottom line in \eqref{bhscale}.} of this Fermi-sea is
\begin{align}
    d_0 =& \sum_{n=0}^{K}\sum_{m=0}^{n+1}\sum_{a=1}^{\dim G}\left(\frac32+n\right)=(N^2-1)\frac{(4K^2+23K+36)(K+1)}{12}
    \cr
    \approx&\,\frac{N^2K^3}{3}~. \end{align}

The leading order correction to the dilation operator can be calculated either by algebraic technics (see appendix
\ref{app-2dim}) or from the Harmonic action. We find that
\begin{equation}
    \dD_2 =-\frac1{2N}\sum_{\substack{k,k',q,q'\\m,m',n,n'}}\mathcal{C}^{(k',m')(q',n')}_{(k,m)(q,n)}
    \tr\cno{\left\{\psi_{(k'|m')},\check\psi_{(k|m)}\right\}\left\{\psi_{(q'|n')},\check\psi_{(q|n)}\right\}}
\end{equation}
Angular momentum conservation law guarantees that $k+q=k'+q'$ and
$m+n=m'+n'$ - otherwise the coefficient vanishes.

When we apply $\dD_2$ to the 2-dim Fermi-sea \eqref{Fermi-sea-2dim},
Pauli's exclusion principle combined with the conservation laws guarantees
that the Fermi-sea is an eigenstate, and the eigenvalue is
\begin{align}
    \dD_2\ket{\Op^{(K)}_{\mathrm{2dim}}}
    =&(N^2-1)\sum_{k=0}^K\sum_{q=0}^K\sum_{m=0}^{k+1}\sum_{n=0}^{q+1}\mathcal{C}^{(q,n)(k,m)}_{(k,m)(q,n)}\ket{\Op^{(K)}_{\mathrm{2dim}}}~. \end{align}
The only coefficient that comes into this sum is
\begin{align}
    \mathcal{C}&^{(q,n)(k,m)}_{(k,m)(q,n)}=
        +\delta_{k=q}\delta_{m=n}\left[h(k)+h(q)\right]+\cr
    &-\Theta(q-k)\Theta(n-m+1)\Theta(q-k-n+m+1)
    \frac{\binom{q-k}{n-m}\binom{k+1}{m}}{\binom{q+1}{n}}\frac{(q+1)}{(q-k)(k+1)}+\cr
    &-\Theta(k-q)\Theta(m-n+1)\theta(k-q-m+n+1)
    \frac{\binom{q+1}{n}\binom{k-q}{m-n}}{\binom{k+1}{m}}\frac{k+1}{(q+1)(k-q)}+\cr
%
    &+
    \frac{\binom{q+1}{n}\binom{k+1}{m}}{\binom{k+q+2}{m+n}}\frac{(k+q+2)}{(q+1)(k+1)}~,
\end{align}
and hence
\begin{align}
    \dD_2\ket{\Op^{(K)}_{\mathrm{2dim}}}
    =~&(N^2-1)\sum_{k=0}^K\Bigg[\sum_{m=0}^{k+1}2h(k)+\cr&
    +\sum_{q=0}^{K}\sum_{m=0}^{k+1}\sum_{n=0}^{q+1}\frac{\binom{q+1}{n}\binom{k+1}{m}}{\binom{k+q+2}{m+n}}\frac{(k+q+2)}{(q+1)(k+1)}+\cr&~
    -2\sum_{q=0}^{K-k-1}\sum_{m=0}^{k+1}\sum_{n=0}^{q+1}\frac{\binom{q+1}{n}\binom{k+1}{m}}{\binom{k+q+2}{m+n}}\frac{(k+q+2)}{(q+1)(k+1)}
    \Bigg]
    \ket{\Op^{(K)}_{\mathrm{2dim}}}=\cr
    =~&(N^2-1)\sum_{k=0}^K\Bigg[2(k+2)h(k)
    +\sum_{q=K-k}^{K}\frac{(k+q+3)(k+q+2)}{(k+1)(q+1)}+\cr&~
    -2\sum_{q=0}^{K-k-1}\frac{(k+q+3)(k+q+2)}{(k+1)(q+1)}
    \Bigg]
    \ket{\Op^{(K)}_{\mathrm{2dim}}}=\cr
    =~&(N^2-1)\frac{(5K+12)(K+1)}{2}~.
\end{align}
Again we see a none-trivial cancelation of leading term of order
$\sim N^2K^2\log K$. We also have a large $K$ relation which is similar to
the one obtained for the 1-dim Fermi-sea
\begin{equation}
    \frac{\dD_2}{d_0}\ket{\Op^{(K)}_{\mathrm{2dim}}}\approx
    \frac{15}{2K}+O(K^{-2})~. \end{equation}

However, $\dD_2$ on these operators receives corrections at order
$g^3$, which are under much less control than what we have discussed
for the 1-dim Fermi-sea \footnote{recall that this operator contains
more types of 'letters'.}. The minimal sector containing
$\psi_{(k|m)}$ is the $\su(1,2|3)$ sector, and hence we need to
consider the full set of partons in this sector
\begin{align}
    \bar Z_{i(n|m)} :=& \frac1{(n)!}(\J_L^-)^m\times\left[(D_{1\dot1})^n\Phi_{4i}\right]
    \cr
    \bar\Psi^i_{(n|m)} :=& \frac1{(n+1)!}(\J_L^-)^m\times\left[(D_{1\dot1})^n\bar\Psi^i_{\dot 1}\right]
    \cr
    \bar F_{(n|m)} :=& \frac1{(n+2)!}(\J_L^-)^m\times\left[(D_{1\dot1})^n\bar F_{\dot 1\dot 1}\right]
\end{align}
By conservation of the left handed angular momentum, the right handed angular momentum, the $\SU(4)$ charges and zero coupling dilatation the only processes that can mix operators are
\begin{align*}
    D_{1\dot1}\cdots D_{2\dot 1}\rightarrow&~ F_{1\dot 1}
    \cr
    D_{1\dot1}\cdots \Psi_{2\;4}\rightarrow&~ \Phi_{4i}\cdots\bar\Psi^{i}_{\dot 1}
    \cr
    D_{2\dot1}\cdots \Psi_{1\;4}\rightarrow&~ \Phi_{4i}\cdots\bar\Psi^{i}_{\dot 1}
    \cr
    \Psi_{1\;4}\cdots \Psi_{2\;4}\rightarrow&~ \Phi_{41}\cdots \Phi_{42}\cdots \Phi_{43}
\end{align*}
where this notation encodes what pairs of 'letters' can be converted
to which single, pair or triplet of 'letters' (at any points in the
expression). All these processes expected \footnote{See the
discussion about the $\su(2|3)\subset\su(1,2|3)$ in
\cite{Beisert:2004ry}} to appear at order $g^3$. The first kind
of mixing will cause two fermions to loose momentum and therefore
will be annihilated by the Pauli's exclusion principle. However all other mixing
are expected to occur and the operator \eqref{Fermi-sea-2dim} is no longer an
eigenstate of $\dD$.

It is an interesting and complicated question to calculate the
$\dD_3$ terms and find an eigenstate which is close to the initial
Fermi-sea. It is even more interesting and more complicated to do so
to all orders. We hope to return to this problem in the future.

\section{Discussion}\label{sec-end}

We studied Fermi-sea operators in $\N=4$ SYM within perturbation
theory up to order $g^4$. We showed that up to this order, the
ground state of the fermionic $\su(1,1)$ sector, and its small
excitations, behave as a Fermi liquid. The correction to the
anomalous dimension of this operator are suppressed by a large
parameter $K$, which is analogous to the Fermi level.

Given that the first few orders of perturbation theory are
parametrically small by this new parameter, $1/K$, it is natural to
ask what is the fate of the Fermi liquid at strong coupling. One
possibility is that the structure found up to $g^4$ is destroyed
allowing $K(\log K)^m$ corrections at order $g^{2n}$ ($m\leq n$). The second
possibility is that the function $F(g^2)$ controlling the leading
$1/K$ correction (defined in \eqref{fgsq}) is diverging at some
finite/infinite value of $g$. The third and most alluring
possibility is that $F(g^2)$ has a finite limit at infinitely strong
coupling.

If the latter option is true than one should be able to trace the
Fermi-sea operator to a classical state (geometry) in AdS. If the
same holds true for operators related to the 1/16-BPS black holes in
$AdS_5$, then it might also suggest that these black holes are not
supersymmetric once $\alpha'$ or $g_s$ corrections are taken into
account, but rather that these corrections are suppressed at large
\sph5 angular momentum ($Q\sim K$).

A possible obstacle to the result are wrapping-like interaction.
These appear in the dilation operator at order where the number of
partons involved is of the size of the operator. In our case wrapping
interaction are suppressed by $g^{2N^2}$, this order cannot be
studied in perturbation theory.

With this caveat in mind, it will be interesting to explore the
following future direction:

$\bullet\qquad$In a recent paper \cite{Zwiebel:2008gr}, which
appeared as the current paper was written, Zwiebel constructs an
iterative algebraic method to calculate the dilatation operator in
the $\psu(1,1|2)$ sector. Using this method it may be possible
\footnote{Assuming the lift to finite N is unique.} (in principle)
to extend our calculation in the fermionic $\su(1,1)$ to higher, and
perhaps all, orders of $g_{ym}$.

$\bullet\qquad$ In condensed matter physics 1-dim Fermi-liquids are
commonly described using bosonization, the bosonized picture allows
for better control over the perturbation theory (in the coupling).
It is possible to repeat the bosonization procedure in our case -
for example, one possible procedure is to define bosonic
rasing/lowering operators \footnote{The continuation of the momentum
to negative values is done formally, and should not effect the
physics near the Fermi level.}
\begin{equation}
    {b^a_{(q)}}^\dag := \frac{i}{\sqrt{q}}\sum_{k=-\infty}^\infty \psi_{(k+q)}^a\check\psi_{(k)}^a~,\spc
    {b^a_{(q)}} := -\frac{i}{\sqrt{q}}\sum_{k=-\infty}^\infty
    \psi_{(k-q)}^a\check\psi_{(k)}^a~~. \end{equation}
This procedure is powerful in condensed matter because in the
bosonic picture the action is quadratic (the bare theory has only a
4 fermion interaction). This will not happen in our case since the
higher orders in $g$ will introduce higher order interaction in the
scalars. However there may be some insight from a bosonic
description.

$\bullet\qquad$ Another interesting issue is to generalize the
construction to the two dimensional Fermi Surface (or small $j_L$) -
in section \ref{sec-2d} we already discussed some difficulties
associated with it. The existence of supersymmetric AdS black holes
with arbitrary $j_L$ strongly motivates the existence of such a generalization but
one probably needs more ingredients on top of the Fermi-sea
discussed here \cite{Berkooz:2006wc}.
Moreover, it could also be that the black holes are not exactly
1/16-BPS. This will make their identification at weak coupling even
more complicated, and it will probably be possible only if they have
features similar to the Fermi-seas that we discussed in this
paper - i.e., the existence of a new small parameter which might
suppress corrections uniformly from weak to strong coupling.

$\bullet\qquad$ Using a Fermi-sea construction could potentially
open the way to building a host of approximately 1/16-BPS operators.
For example, if we add the boson which completes the sector to the
$\su(1,1|1)$ sector, is easy to find additional ways to make $\dD_2$
parametrically small in the presence of the Fermi-sea (although it
is not clear how to control these in higher orders in $g_{ym}$).
More specifically we add the partons
\begin{equation}
    \phi_{(n)}:=\frac1{n!}(D_{1\dot1})^n\Phi_{14}~, \end{equation}
the order $g^2$ anomalous dimension is
\begin{align}
    \dD_{2} =& \dD_2^{(fermions)} +\cr&
    +\frac{1}{N}\sum_{n,m,k,q=0}^\infty\left(\frac{\Theta(m-q)}{n+1}-\frac{\Theta(n-q)}{n-q}\right)\tr\cno{\left[\phi_{(k)},\check\phi_{(m)}\right]\left\{\psi_{(q)},\check\psi_{(n)}\right\}}\delta_{n+m=k+q}
    +\cr&
    +\frac{1}{N}\sum_{n,m,k,q=0}^\infty\frac{\Theta(n-k+1)}{n+1}\tr\cno{\left[\phi_{(k)},\psi_{(q)}\right]\left[\check\phi_{(m)},\check\psi_{(n)}\right]}\delta_{n+m=k+q}
    +\cr&
    -\frac1{N}\sum_{n,m,k,q=0}^\infty\frac{\Theta(n-q)}{n-q}\tr\cno{\left[\phi_{(k)},\check\phi_{(m)}\right]\left[\phi_{(q)},\check\phi_{(n)}\right]}\delta_{n+m=k+q}
    +\cr&
    -\frac{2}{N}\sum_{m=0}^\infty h(m)\tr\cno{\phi_{(m)}\check\phi_{(m)}}~. \end{align}

An almost 1/16-BPS operator can be built by multiplying the
Fermi-sea with a bosonic operator
\begin{equation}
    \Op = \Op^{(K)}_{\mathrm{1dim}}\tr\left(\phi_{(0)}^M\right).
\end{equation}
The operator is an eigenstate of the dilatation operator (to all orders in perturbation theory) with anomalous dimension at order $g^2$:
\begin{equation}
    \dD_2\ket{\Op}= \left[(N^2-1)2(K+1)-2Mh(K+1)\right]\ket{\Op}~. \end{equation}
By solving $\dD_2=0$ we find that for a large scalar contribution
\begin{equation}
    M =(N^2-1)\frac{(K+1)}{h(K+1)}\sim N^2 K/\log K~.
\end{equation}
Thus we can get as close to a BPS operator (since $M$ must be an integer
we cannot have an BPS state).

In the case that the large AdS black holes of \cite{Gutowski:2004ez}
are indeed exactly supersymmetric (with $\alpha'$ and $g_s$
corrections included) we expect a large amount (of order $N^2$) of
1/16-BPS semi-short multiplets. The above example demonstrates how a
Fermi-sea facilitates a cancelation of anomalous dimension (due to
fermion-boson interaction). A detailed search for 1/16-BPS operator
following these idea and techniques is a promising but challenging
avenue.

$\bullet\qquad$ Finally, it will be interesting to explore these
techniques for application to condensed matter systems. Fermi-sea
constructions are of major importance in many strongly coupled
condensed matter system, but so far they were not part of any
AdS/CFT model. If from $N=4$ SYM we can learn about the description
of Fermi-seas in $AdS_5$, we may be able to explore phenomena
related to Fermi-sea in lower dimension $AdS_d$ spaces ($d=4,3,2$)
as well. These in turn may be important as models for many
interesting condensed matter system (superconductors, quantum hall
effect etc.)

\acknowledgments

We are grateful to O.~Aharony, Z.~Komargodski, D.A.~Kosower,
S.~Minwalla, B.~Pioline, J.~Simon, and K.~Zarembo for valuable
discussions. D.R. would like to give special thanks for R.~Ilan and
A.~Stern who make a special effort to teach condensed matter physics
to the barbarians.

This work was supported by the Israel-U.S. Binational Science
Foundation, by a center of excellence supported by the Israel
Science Foundation (grant number 1468/06), by a grant (DIP H52) of
the German Israel Project Cooperation, by the European network
MRTNCT- 2004-512194, and by a grant from G.I.F., the German-Israeli
Foundation for Scientific Research and Development, and by the
Einstein-Minerva center for theoretical physics.

\appendix
\section{The $\psu(2,2|4)$ algebra and the partons of $\N=4$ SYM}\label{app-par}

In this appendix we review some properties of the $\psu(2,2|4)$ algebra and it's representations. More details can be found in \cite{Beisert:2004ry}. The generators of the $\psu(2,2|4)$ algebra are
\begin{itemize}
   \item The compact bosonic $\su(2)_L\times\su(2)_R\times\su(4)$ generators ${\mathfrak{L}^\alpha}_\beta$ , $\mathfrak{\bar L}^{\dot\alpha}_{~\dot\beta}$ , ${\mathfrak{R}^a}_b$.
   \item The non-compact bosonic translation, dilatation and special conformal generators $\mathfrak{P}_{\alpha\dot\alpha}$ , $\mathfrak{D}$ , $\mathfrak{K}^{\dot\alpha\alpha}$.
   \item The supercharges $\Q^a_{\alpha}$ , $\bar \Q_{\dot\alpha a}$ and super-conformal supercharges $\mathfrak{S}_a^{\alpha}$ , $\mathfrak{\bar S}^{\dot\alpha a}$~. \end{itemize}
The algebra can be extended to $\mathfrak{u}(2,2|4)$ by introducing
two additional $\mathrm{U}(1)$ generators (commuting with the
$\psu(2,2|4)$) - the central charge $\mathfrak{C}$ and the
hypercharge $\mathfrak{B}$. Physical states must satisfy
$\mathfrak{C}=0$. We use the oscillator representation to write the
(zero coupling) generators, and then all commutation relations can
be derived from the commutation relation of the oscillators
\eqref{oscilator}
\begin{align}
    {\mathfrak{L}^\alpha}_\beta =\,& \ab_\beta^\dag\ab^\alpha-\frac12\delta_\beta^\alpha\,\ab_\gamma^\dag\ab^\gamma
    &
    \mathfrak{B} =\,& \frac12\ab_\gamma^\dag\ab^\gamma-\frac12\bb_{\dot\gamma}^\dag\bb^{\dot\gamma}
\cr
    \mathfrak{\bar L}^{\dot\alpha}_{~\dot\beta} =&
    \bb_{\dot\beta}^\dag\bb^{\dot\alpha}-\frac12\delta_{\dot\beta}^{\dot\alpha}\,\bb_{\dot\gamma}^\dag\bb^{\dot\gamma}
    &
    \mathfrak{D} =\,& 1+\ab_\gamma^\dag\ab^\gamma+\bb_{\dot\gamma}^\dag\bb^{\dot\gamma}
\cr
    {\mathfrak{R}^a}_b =\,& \cb_b^\dag\cb^a-\frac12\delta_b^a\,\cb_c^\dag\cb^c
    &
    \mathfrak{C} =\,& 1-\frac12\ab_\gamma^\dag\ab^\gamma+\frac12\bb_{\dot\gamma}^\dag\bb^{\dot\gamma}-\frac12\cb_c^\dag\cb^c
\cr
    \Q^a_\alpha =\,& \ab_\alpha^\dag\cb^a
    &
    \bar\Q_{\dot\alpha a}=\,& \bb_{\dot\alpha}^\dag\cb_a^\dag
\cr
    \mathfrak{S}_a^\alpha =\,& \cb_a^\dag\ab^\alpha
    &
    \mathfrak{\bar S}^{\dot\alpha a}=\,& \bb^{\dot\alpha}\cb^a
\cr
    \mathfrak{P}_{\alpha\dot\beta}=\,&\ab_\alpha^\dag\bb_{\dot\beta}^\dag
    &
    \mathfrak{K}_{\alpha\dot\beta}=\,&\ab^\alpha\bb^{\dot\beta}~.
\end{align}
In the oscillator language representations of $\psu(2,2|4)$ are
labeled by a set of weights\footnote{We use Fraktur font for
generators and uncapitalized Latin font for weights}
$w=\left(d_0;j_L,j_R;q_1,p,q_2;B,L\right)$ which are the charges of
the dilatation (zero coupling), left and right angular momenta,
$\SU(4)$ weights, hypercharge, and spin-chain length. The weights
are related to the Cartan generators
\begin{align}
    \mathfrak{q}_1 =\,&{\R^2}_2-{\R^1}_1
    &
    \mathfrak{p} =\,&{\R^2}_2-{\R^2}_2
    &
    \mathfrak{q}_2 =\,&{\R^4}_4-{\R^3}_3
    \cr
    \J_L^3 =\,&\frac12{\mathfrak{L}^1}_1-\frac12{\mathfrak{L}^2}_2
    &
    \J_R^3 =\,&\frac12\mathfrak{\bar L}^{\dot 1}_{~\dot 1}-\frac12\mathfrak{\bar L}^{\dot 2}_{~\dot
    2}\ \ \ \ .
    &
\end{align}
We define two combinations of Cartan generators
\begin{align}\label{deltot}
    \Delta_1 =& \D-2\J^3_L-\frac12\mathfrak{q}_1-\mathfrak{p}-\frac32\mathfrak{q}_2
    \cr
    \Delta_2 =& \D-2\J^3_R-\frac32\mathfrak{q}_1-\mathfrak{p}-\frac12\mathfrak{q}_2
\end{align}
The unitarity bounds of $\N=4$ SYM are $\Delta_1\geq0$ and
$\Delta_1\geq0$. For primary operators one should subtract
$c_m=0,1,2$ form the RHS of \eqref{deltot}, for 0,1 or 2 non-singlet
angular momentum representations. We conclude with a detailed list
of the weights of all partons, we set all lorentz indices to the
highest-weight.

\smallskip
\begin{tabular}{||c|c|c|c|c|c|c||}
\hline\hline
                 & $d_0$ & $(2j_L, 2j_R)$ & $(q_1,p,q_2)$        & $B$ & $\Delta_1$ & $\Delta_2$  \\
\hline
 $(D_{1\dot1})^{n}\Psi_{1\;4}$ & $n+3/2$ &  $(n+1,n)$ & $(0,0,1)$     & $1/2$ &   $0$  &  $0$    \\
 $(D_{1\dot1})^{n}\Phi_{34}$ & $n+1$   &  $(n,n)$   & $(0,1,0)$  & $0$ &   $0$  &  $0$   \\
 $(D_{1\dot1})^{n}\Phi_{24}$ & $n+1$   &  $(n,n)$   & $(1,-1,1)$ & $0$ &   $0$  &  $0$   \\
 $(D_{1\dot1})^{n}\bar\Psi^1_{\dot1}$ & $n+3/2$ &  $(n,n+1)$  & $(1,0,0)$ & $-1/2$ &  $0$ & $0$  \\
\hline
   $(D_{1\dot1})^{n}F_{11}$    & $n+2$   &  $(n+2,n)$ & $(0,0,0)$  & $1$ & $0$  &  $2$        \\
 $(D_{1\dot1})^{n}\Psi_{1\;3}$ & $n+3/2$ &  $(n+1,n)$ & $(0,1,-1)$ & $1/2$ & $0$  &  $2$      \\
 $(D_{1\dot1})^{n}\Psi_{1\;2}$ & $n+3/2$ &  $(n+1,n)$ & $(1,-1,0)$ & $1/2$ & $0$  &  $2$      \\
 $(D_{1\dot1})^{n}\Psi_{1\;1}$ & $n+3/2$ &  $(n+1,n)$ & $(-1,0,0)$ & $1/2$ & $2$  &  $2$      \\
 $(D_{1\dot1})^{n}\Phi_{23}$ & $n+1$   &  $(n,n)$   & $(1,0,-1)$ & $0$ & $0$  &  $2$     \\
 $(D_{1\dot1})^{n}\Phi_{14}$ & $n+1$   &  $(n,n)$   & $(-1,0,1)$ & $0$ & $2$  &  $0$     \\
 $(D_{1\dot1})^{n}\Phi_{13}$ & $n+1$   &  $(n,n)$   & $(-1,1,-1)$& $0$ & $2$  &  $2$     \\
 $(D_{1\dot1})^{n}\Phi_{12}$ & $n+1$   &  $(n,n)$   & $(0,-1,0)$ & $0$ & $2$  &  $2$     \\
 $(D_{1\dot1})^{n}\bar\Psi^2_{\dot1}$ & $n+3/2$ &  $(n,n+1)$  & $(-1,1,0)$ & $-1/2$ & $2$ & $0$  \\
 $(D_{1\dot1})^{n}\bar\Psi^3_{\dot1}$ & $n+3/2$ &  $(n,n+1)$  & $(0,-1,1)$ & $-1/2$ & $2$ & $0$  \\
 $(D_{1\dot1})^{n}\bar\Psi^4_{\dot1}$ & $n+3/2$ &  $(n,n+1)$  & $(0,0,-1)$ & $-1/2$ & $2$ & $2$  \\
 $(D_{1\dot1})^{n}\bar F_{\dot1\dot1}$    &  $n+2$   &  $(n,n+2)$  & $(0,0,0)$  & $-2$ & $2$ &  $0$   \\
\hline\hline
\end{tabular}

\section{Gauge transformation for finite $N$ operators}\label{app-gauge}

The usage of gauge transformation to generate identities between operators in the finite $N$ language is explained in \cite{Beisert:2004ry}. We start from the generator of gauge transformations \footnote{$C$ runs over all partons, i.e all fields with derivatives such that the lorentz indices are multiplied to form the highest possible weight.}
\begin{equation}
    \mathfrak{j}=i\sum_C\cno{\left[W_C,\check W^C\right\}}~. \end{equation}
By plugging $\mathfrak{j}$ into operational expression one can generate a large set of gauge identities. One identity we use in the main text \eqref{newgg}, is proved as follows
\begin{align}
    0\widehat{=}\,&-i \tr\mathfrak{j}\cno{\left[W_A,\check W^B\right\}}
    =\cr=&\sum_{C}\tr\cno{\left[W_C,\check W^C\right]\left[W_A,\check W^B]\right\}}
    +2N\tr\cno{W_A\check W^B}~, \end{align}
by restriction to the fermionic $\mathfrak{su}(1,1)$ sector we find
\begin{equation}
    \tr\cno{\psi_{(m)}\check\psi_{(m)}}
    =
    -\frac1{2N}
    \sum_{n=0}^\infty
    \tr\cno{\left\{\psi_{(m)},\check\psi_{(m)}\right\}\left\{\psi_{(n)},\check\psi_{(n)}\right\}}~.
\end{equation}
The second identity we need \eqref{gid2} is proved using
\begin{align}
    0\widehat{=}\,&-i \tr\mathfrak{j}\cno{\left[\left[W_A,W_B\right\},\check W^C\right\}}
    =\cr
    =&
    \sum_{D}\tr\cno{\left[W_D,\check W^D\right\}\left[\left[W_A,W_B\right\},\check W^C\right\}}
    +2N\tr\cno{W_A\left[W_B,\check W^C\right\}}~,
\end{align}
restricting to the fermionic $\mathfrak{su}(1,1)$ sector we find
\begin{multline}
    \tr\cno{\left\{\psi_{(k)},\psi_{(q)}\right\}\check\psi_{(k+q+1)}}
    \widehat{=}\\
    \widehat{=}\,
    -\frac1{2N}\sum_{m=0}^\infty\tr\cno{\left\{\psi_{(m)},\check\psi_{(m)}\right\}
    \left[\left\{\psi_{(q)},\psi_{(k)}\right\},\check\psi_{(k+q+1)}\right]}~.
\end{multline}

\section{Zwiebel's solution for the $g^4$ dilatation operator}\label{app-2loop}

An algebraic solution for the order $g^4$ dilatation operator was found by Zwiebel \cite{Zwiebel:2005er}. The solution is based on the $\psu(1,1|2)$ sector. The sector is generated by the partons
\begin{align}
    &\psi_{(n)} := \frac1{(n+1)!}\left(D_{1\dot1}\right)^n\Psi_{1\,4}
    \cr
    &\bar\psi_{(n)} := \frac1{(n+1)!}\left(D_{1\dot1}\right)^n\bar\Psi^1_{\dot 1}
    \cr
    &\phi^2_{(n)} := \frac1{(n)!}\left(D_{1\dot1}\right)^n\Phi_{24}
    \cr
    &\phi^3_{(n)} := \frac1{(n)!}\left(D_{1\dot1}\right)^n\Phi_{34}~.
\end{align}
States generated by these partons sit in representation of a $\psu(1,1|2)\times\psu(1|1)^2$ subalgebra of $\psu(2,2|4)$ \begin{itemize}
\begin{subequations}
  \item the $\mathfrak{su}(1,1|2)$ symmetry is generated by
\begin{align}
    \J^0(g)=&\, -\mathcal{L}+2\mathfrak{D}_0+\dD(g) &
    \R^0 =&\, \R^2_2-\R^3_3
    \cr
    \J^{++}(g) =&\, \mathfrak{P}_{11}(g) &
    \J^{--}(g) =&\, \mathfrak{K}^{11}(g)
    \cr
    \R^{22}(g) =&\, \R^3_2&
    \R^{33}(g) =&\, \R^2_3\cr
    \mathfrak{Q}^{+i}(g)=&\,\mathfrak{Q}^i_1(g)&
    \mathfrak{\bar Q}^{+i}(g)=&\,\mathfrak{\bar Q}_{1i}(g)&
    \cr
    \mathfrak{Q}^{-i}(g)=&\,\mathfrak{\bar S}^{1i}(g)&
    \mathfrak{\bar Q}^{-i}(g)=&\,\mathfrak{S}^1_{i}(g)&
\end{align}
where $i=2,3$ and $\R^a_b$ are generators of the $\SU(4)$ symmetry.
  \item The $\mathfrak{psu}(1|1)^2$ symmetry is generated by
\begin{align}
    \T^+(g)=&\,\mathfrak{\bar Q}_{24}(g)&
    \T^-(g)=&\,\mathfrak{S}^{2}_{1}(g)
    \cr
    \bar\T^+(g)=&\,\mathfrak{Q}^{1}_{2}(g)&
    \bar\T^-(g)=&\,\mathfrak{\bar S}^{24}(g)
    \cr
    \dD(g)&&
    \mathcal{L}
\end{align}
\end{subequations}
\end{itemize}
With the relation
\begin{equation}
    \dD(g)=2\left\{\T^+(g),\bar\T^-(g)\right\}=2\left\{\T^-(g),\bar\T^+(g)\right\}~.
\end{equation}
The algebra includes the dilatation operator, thus the sector is
closed under mixing. Similar to the $\su(1,1)$ sector we use
explicit relations between charges in the $\psu(1,1|2)$ sector. A
reduction of the above to the fermionic $\su(1,1)$ is carried out by
considering states that consists only of $\psi_{(n)}$ parton.

The first step in Zwiebel's solution is the $g^1$ corrections to the
$\psu(1|1)^2$ supercharges $\T^\pm$ and $\bar\T^\pm$, which we
already discussed in the main body of the text \eqref{supercharges}.
Next Zwiebel defines two new operators,
\begin{align}
    \mathfrak{h} :=& \sum_{n=0}^\infty
    \frac12h(n+1)\tr\cno{\psi_{(n)}\check \psi_{(n)}}
    +\sum_{n=0}^\infty \frac12h(n+1)\tr\cno{\bar\psi_{(n)}\check{\bar\psi}_{(n)}}
    \cr
    &+\sum_{n=0}^\infty \frac12h(n)\sum_{i=2}^3\tr\cno{\phi^i_{(n)}\check \phi^i_{(n)}}
\cr
    \mathfrak{r} :
    =& \left\{\T^-_1,\left[\bar\T^+_1,\mathfrak{h}\right]\right\}
    -\left\{\T^+_1,\left[\bar\T^-_1,\mathfrak{h}\right]\right\}~. \end{align}
Using these two operators Zwiebel shows that the $g^2$ correction to
the $\psu(1,1|2)$ algebra are
\begin{align}
    \mathfrak{R}^0_2=&\,0
&
    \J^0_2=&\,\dD_2
&
    \mathfrak{X}_2^\pm =&\,\pm \left[\mathfrak{X}_0^\pm,\mathfrak{r}\right]+\left[\mathfrak{X}_0^\pm,\mathfrak{y}\right]~,
\end{align}
with $\mathfrak{X}^\pm\in\left\{\J^{++},\J^{--},\Q^{\pm
i},\bar\Q^{\pm i}\right\}$. The $g^3$ corrections to the
$\psu(1|1)^2$ supercharges are
\begin{align}
    \T_3^\pm =&\, \pm\left[\T^\pm_1,\mathfrak{r}\right]
    +\left[\T^\pm_1,\mathfrak{y}\right]+\alpha\T^\pm_1
    &
    \bar\T_3^\pm =&\, \pm\left[\bar\T^\pm_1,\mathfrak{r}\right]
    +\left[\bar\T^\pm_1,\mathfrak{y}\right]+\alpha\T^\pm_1~.
\end{align}
In the above, the parameter $\alpha$ is related to a coupling
redefinition $g\mapsto g+\alpha g^3$ and $\mathfrak{h}$ to a
similarity transformation. It is possible to choose a regularization
scheme such that both are zero. Finally the $g^4$ correction to
$\dD$ is found by anti-commutator of $\T^+$ and $\bar\T^-$ resulting in
\begin{align}\label{deltaD4-ap}
    \dD_4= & 2\left\{\T^+(g),\bar\T^-(g)\right\}_4 =\cr =&2\left\{\T^+_3,\bar\T^-_1\right\}+2\left\{\T^+_1,\bar\T^-_3\right\}=\cr
    =&2\left\{\bar\T^-_1,\left[\T^+_1,\left\{\T^-_1,\left[\bar\T^+_1,\mathfrak{h}\right]\right\}\right]\right\}
    +2\left\{\T^+_1,\left[\bar\T^-_1,\left\{\bar\T^+_1,\left[\T^-_1,\mathfrak{h}\right]\right\}\right]\right\}
\end{align}
In the above we already removed the coupling redefinition and similarity transformation. \footnote{In general there is a redefinition of generators such that ~$\dD_4\mapsto\dD_4 + 2\alpha\dD_2 + \left[\dD_2,\mathfrak{h}\right]$.}

\section{The anomalous dimension of the 1-dim Fermi-sea}\label{app-g4-Fermi}

In this appendix we evaluate \eqref{deltaD4-operator} on the 1-dim Fermi-sea state \eqref{Fermi-sea-1dim}. First we take care of the operational structures appearing the $\dD_4$. The key feature is that Pauli's exclusion principle combined with angular momentum conservation guarantees that all creation operators ($\psi_{(p)}$) are contracted by annihilation operators ($\check\psi_{(p)}$) generating delta-functions in momentum space and in the gauge group indices. The fact that the contractions are done only after the creation operator hits the Fermi-sea state results in $\Theta$-functions limiting the summation range. The outcomes of this feature are summarized by the following identities :
\begin{subequations}
\begin{equation}
    \tr\cno{\psi_{(m)}\check\psi_{(m)}}\ket{\Op^{(K)}_{\mathrm{1dim}}}
    =
    +\Theta(K+1-m)(N^2-1)\ket{\Op^{(K)}_{\mathrm{1dim}}}
\end{equation}
\begin{multline}
    \tr\cno{\left\{\psi_{(k)},\check\psi_{(m)}\right\}\left\{\psi_{(q)},\check\psi_{(n)}\right\}}\ket{\Op^{(K)}_{\mathrm{1dim}}}
    =\\=
    -2N(N^2-1)\delta_{q=m}\delta_{k=n}\Theta(K+1-m)\Theta(K+1-n)\ket{\Op^{(K)}_{\mathrm{1dim}}}
\end{multline}
\begin{multline}
    \tr\cno{\psi_{(q)}\check\psi_{(n)}\psi_{(k)}\check\psi_{(m)}}\ket{\Op^{(K)}_{\mathrm{1dim}}}
    =\Theta(K+1-m)\Theta(K+1-n)
    \times\\
    \times\frac{(N^2-1)^2}{N}
    \left(\delta_{q=n}\delta_{k=m}-\delta_{q=m}\delta_{k=n}\right)
    \ket{\Op^{(K)}_{\mathrm{1dim}}}
\end{multline}
\begin{multline}
    \tr\cno{\psi_{(q)}\check\psi_{(n)}}\tr\cno{\psi_{(k)}\check\psi_{(m)}}\ket{\Op^{(K)}_{\mathrm{1dim}}}
    =\Theta(K+1-m)\Theta(K+1-n)
    \times\\
    \times(N^2-1)^2
    \left(\delta_{q=n}\delta_{k=m}-\frac1{N^2-1}\delta_{q=m}\delta_{k=n}\right)
    (\ket{\Op^{(K)}_{\mathrm{1dim}}}
\end{multline}
\begin{multline}
    \tr\cno{\psi_{(q)}\psi_{(k)}}\tr\cno{\check\psi_{m)}\check\psi_{(n)}}\ket{\Op^{(K)}_{\mathrm{1dim}}}
    =\Theta(K+1-m)\Theta(K+1-n)
    \times\\
    \times(N^2-1)
    \left(\delta_{q=n}\delta_{k=m}-\delta_{q=m}\delta_{k=n}\right)
    \ket{\Op^{(K)}_{\mathrm{1dim}}}
\end{multline}
\begin{align}
    \tr\cno{&\left[\left\{\psi_{(r)},\check\psi_{(m)}\right\},\check\psi_{(l)}\right]
    \left[\left\{\psi_{(s)},\check\psi_{(n)}\right\}\psi_{(t)}\right]}\ket{\Op^{(K)}_{\mathrm{1dim}}}
    =\cr
    =&\Theta(K+1-m)\Theta(K+1-n)\Theta(K+1-l)\,4N^2(N^2-1)
    \times\cr
    &\times
    \left(-\delta_{m=s}\delta_{n=r}\delta_{l=t}
    +\delta_{m=s}\delta_{n=t}\delta_{l=r}
    +\frac12\delta_{m=t}\delta_{n=r}\delta_{l=s}\right)
    \ket{\Op^{(K)}_{\mathrm{1dim}}}
\end{align}
\end{subequations}
Combining the above result with the coefficients in \eqref{deltaD4-operator} we express that $g^4$ anomalous dimension of the Fermi-sea by a set of sums:
\begin{align}
    \dD_4 = (N^2-1)&\sum_{r=0}^K\sum_{s=K-r}^{K}\Bigg[
    \mathcal{C}_{(n,m)}\xi_{(n,m)}-\frac12\bar\mu_{(n,m)}-\frac12\mathcal{C}_{(n,m)}\bar\nu_{(n,m)}
    +\cr&
    +\sum_{q=0}^K\left(C_{(r,s)}\nu_{[(r;r+s+1),(s;q)]}
    -\frac12C_{(r,s)}\nu_{[(q;r+s+1),(s;q)]}\right)
    +\cr&
    +\sum_{q=0}^K\left(\mu_{[(r+s+1;r),(q;s)]}
    -\frac12\mu_{[(r+s+1;q),(q;s)]}\right)
    \Bigg]
\end{align}

Note that we did not applied any large $N$ limit, the non-planar contributions (the double trace part of $\dD_4$) has the same order of magnitude in $N$ as the planar contributions. We were unable to calculate the expression above analytically, instead we use a numerical calculation. Even thou the expression is quite complicated the series is easily found numerically
\begin{equation}
    \dD_4\ket{\Op^{(K)}_{\mathrm{1dim}}}=-(N^2-1)(2K+2)\ket{\Op^{(K)}_{\mathrm{1dim}}}~.
\end{equation}
We checked this result for all values of $K$ up to $K=150$, and found a perfect match.

\section{The $g^2$ anomalous dimension of the 2-dim Fermi-sea}\label{app-2dim}

The fermions $\psi_{(k|m)}$ transform in a representation of $\su(1,2)$, up to order $g^2$ this is a closed sector however at order $g^2$ they mix within the larger $\su(1,2|3)$ sector. We will focus on the $g^2$ correction to the dilatations operator allowing us to focus on the smaller symmetry.  The algebra which closes on the fermions is $\mathfrak{su}(1,2)\times\mathfrak{u}(1|1)$.
\begin{itemize}
    \item The $\mathfrak{su}(1,2)$ algebra
    \begin{subequations}
    \begin{align}
        &\mathfrak{J'}_{++}(g)=\mathfrak{K}^{2\dot1}(g)&
        &\mathfrak{J'}_{-+}(g)=\mathfrak{P}_{\dot11}(g)&\cr
        &\mathfrak{J'}_{+-}(g)=\mathfrak{K}^{1\dot1}(g)&
        &\mathfrak{J'}_{--}(g)=\mathfrak{P}_{\dot12}(g)&\cr
        &\mathfrak{J'}_{0}(g)=-\mathfrak{D}-\frac12\left(\mathfrak{\bar L}^1_1-\mathfrak{\bar L}^2_2\right)&
        &\mathfrak{L'}_{+} = \mathfrak{L}^{2}_{1}\cr
        &\mathfrak{L'}_{-} = \mathfrak{L}^{1}_{2}&
        &\mathfrak{L'}_{0} = \frac12\left(\mathfrak{L}^{1}_{1}-\mathfrak{L}^{2}_{2}\right)
    \end{align}
    \item The $\mathfrak{u}(1|1)$ algebra
    \begin{align}
    &\mathcal{L}&
    &\mathfrak{Q'}_-(g)=\mathfrak{\bar Q}_{\dot 2 4}(g)&
    &\mathfrak{Q'}_+(g)=\mathfrak{\bar S}^{\dot 2 4}(g)&
    &\mathfrak{\delta D}(g)&
    \end{align}
    The commutation relation (inherited from the $\mathfrak{psu}(2,2|4)$ algebra) are
    \begin{align}
        &\left[\mathcal{L},\mathfrak{Q'}_\pm\right]=\mp\mathfrak{Q'}_\pm&
        &\left\{\mathfrak{Q'}_+,\mathfrak{Q'}_-\right\}=\frac12\dD
        &\left[\mathcal{L},\dD\right]=0 &
        &\left[\dD,\mathfrak{Q'}_\pm\right]=0 &
    \end{align}
    \end{subequations}
\end{itemize}
The $\mathfrak{su}(1,2)$ and $\mathfrak{u}(1|1)$ generators commute with each other.
At the leading order in $g$, the generator of $\mathfrak{su}(1,2)$ are the free theory ($g^0$) generators. For the $\mathfrak{u}(1|1)$ the supercharges are of order $g^1$ and $\dD$ is of order $g^2$.
The operation of $\mathfrak{su}(1,2)$ on the spin states is
\begin{align}
    \mathfrak{J'}_{++}\ket{(k,m)} =&~k\,\delta_{m\neq0}\,\ket{(k-1,m-1)}\cr
    \mathfrak{J'}_{--}\ket{(k,m)} =&~(m+1)\ket{(k+1,m+1)}\cr
    \mathfrak{L'}_{+}\ket{(k,m)}  =&~(k-m+2)\,\delta_{m\neq 0}\,\ket{(k,m-1)}\cr
    \mathfrak{L'}_{-}\ket{(k,m)}  =&~(m+1)\,\delta_{k-m+1\neq 0}\,\ket{(k,m+1)}~.
\end{align}
For the supercharge we make the ansatz (most general form that conserves the charges)
\begin{align}\label{ansatz}
    &\mathfrak{Q'}_-\ket{(k,m)}=\sum_{q=0}^{k-1}\sum_{\substack{n=\\\max(0,m+q-k)}}^{\min(q+1,m)}c^-_{q,n;k,m}\ket{(q,n)(k-q-1,m-n)}
    \cr
    &\mathfrak{Q'}_+\ket{(q,n)(k-q,m-n)}=c^+_{q,n;k,m}\ket{(k+1,m)}~.
\end{align}
Next we prove that the algebra forces that
\begin{equation*}
    c^{-}_{q,n;k,m}=a_-~,
\end{equation*}
up to some unknown constant $a_-$. The key ingredient is that the $\mathfrak{u}(1|1)$  and $\mathfrak{su}(1,2)$ factors must commute also in perturbation theory (non zero coupling) however the commutation is up to a term that vanishes due to the gauge structure. We need to check that the ansatz \eqref{ansatz} is consistent with the algebra, i.e the supercharges commutes with the generators of $\mathfrak{su}(1,2)$ up to gauge transformations. \footnote{More correctly, any operator with the correct charges that vanishes when the trace is applied to the spin chain can appear in the commutation of the supercharges and $\mathfrak{su}(1,2)$.} Both supercharges conserves $\mathfrak{J'}_0$ and $\mathfrak{L'}_0$ trivially. For the other generators we find
\begin{subequations}
\begin{align}
    \left[\mathfrak{Q'}_-,\mathfrak{L'}_{+}\right]\ket{(k,m)}=\hspace{-5em}&\cr
    =~&\sum_{q=0}^{k-1}\sum_{\substack{n=\\\max(0,m+q-k-1)}}^{\min(q+1,m-1)}\biggl[
    \delta_{m\neq{0}}(k-m+2)c^-_{q,n;k,m-1}
    +\cr&
    -c^-_{q,n+1;k,m}(q-n+1)
    +\cr&
    -c^-_{q,n;k,m}(k-q-m+n+1)\biggr]\ket{(q,n)(k-q-1,m-n-1)}
\\
   \left[\mathfrak{Q'}_-,\mathfrak{L'}_{-}\right]\ket{(k,m)}=\hspace{-5em}&\cr
    =~&\sum_{q=0}^{k-1}\sum_{\substack{n=\\\max(0,m+q-k+1)}}^{\min(q+1,m+1)}\biggl[
    c^-_{q,n;k,m+1}\delta_{k-m+1\neq{0}}(m+1)
    -c^-_{q,n-1;k,m}n
    +\cr&
    -c^-_{q,n;k,m}(m-n+1)\biggr]\ket{(q,n)(k-q-1,m-n+1)}
\\
    \left[\mathfrak{Q'}_-,\mathfrak{J'}_{-+}\right]\ket{(k,m)}=\hspace{-5em}&\cr
    =~&\sum_{q=0}^{k}\sum_{\substack{n=\\\max(0,m+q-k-1)}}^{\min(q+1,m)}\biggl[
    (k+2-m)c^-_{q,n;k+1,m}
    -c^-_{q-1,n;k,m}\delta_{q\neq{0}}(q-n+1)
    +\cr&
    -c^-_{q,n;k,m}\delta_{q\neq{k}}(k-q-m+n+1)\biggr]\ket{(q,n)(k-q,m-n)}
\\
    \left[\mathfrak{Q'}_-,\mathfrak{J'}_{+-}\right]\ket{(k,m)}=\hspace{-5em}&\cr
    =~&\sum_{q=0}^{k-2}\sum_{\substack{n=\\\max(0,m+q-k+1)}}^{\min(q+1,m)}\biggl[
    \delta_{k-m+1\neq{0}}kc^-_{q,n;k-1,m}
    -c^-_{q+1,n;k,m}(q+1)
    +\cr&
    -c^-_{q,n;k,m}(k-q-1))\biggr]\ket{(q,n)(k-q-2,m-n)}
\\
    \left[\mathfrak{Q'}_-,\mathfrak{J'}_{--}\right]\ket{(k,m)}=\hspace{-5em}&\cr
    =~&\sum_{q=0}^{k}\sum_{\substack{n=\\\max(0,m+q-k)}}^{\min(q+1,m+1)}\biggl[
    (m+1)c^-_{q,n;k+1,m+1}
    -c^-_{q-1,n-1;k,m}\delta_{q\neq{0}}n
    +\cr&
    -c^-_{q,n;k,m}\delta_{q\neq{k}}(m-n+1)\biggr]\ket{(q,n)(k-q,m-n+1)}
\\
    \left[\mathfrak{Q'}_-,\mathfrak{J'}_{++}\right]\ket{(k,m)}=\hspace{-5em}&\cr
    =~&\sum_{q=0}^{k-2}\sum_{\substack{n=\\\max(0,m+q-k)}}^{\min(q+1,m-1)}\biggl[
    \delta_{m\neq{0}}kc^-_{q,n;k-1,m-1}
    -c^-_{q+1,n+1;k,m}(q+1)
    +\cr&
    -c^-_{q,n;k,m}(k-q-1)\biggr]\ket{(q,n)(k-q-2,m-n-1)}~.
\end{align}
\end{subequations}
Demanding the the RHS vanishes we find that $c^-_{q,n;k,m}=a_-$. Indeed the commutators vanish up to gauge transformations
\begin{subequations}
\begin{align}
    &\left[\mathfrak{Q'}_-,\mathfrak{L'}_{+}\right]\ket{(k,m)}=0
\\
    &\left[\mathfrak{Q'}_-,\mathfrak{L'}_{-}\right]\ket{(k,m)}=0
\\
    &\left[\mathfrak{Q'}_-,\mathfrak{J'}_{-+}\right]\ket{(k,m)}=
    a_-\Bigl(\ket{(0,0)(k,m)}+\ket{(k,m)(0,0)}\Bigr)
\\
    &\left[\mathfrak{Q'}_-,\mathfrak{J'}_{+-}\right]\ket{(k,m)}=0
\\
    &\left[\mathfrak{Q'}_-,\mathfrak{J'}_{--}\right]\ket{(k,m)}=
    \delta_{k\neq0}a_-\Bigl(\ket{(0,1)(k,m)}+\ket{(k,m)(0,1)}\Bigr)
\\
    &\left[\mathfrak{Q'}_-,\mathfrak{J'}_{++}\right]\ket{(k,m)}=0~.
\end{align}
\end{subequations}
Note: One may start with the most general gauge transformation on the right hand side of the commutators. Using the consistency of the algebra and the physical demand that a single fermion (with no derivative) cannot split due to the supercharge.\footnote{Unlike $\psi_{k|m}$, a single $\Psi_{14}$ without derivative is not a composite operator and there is no renormalization that can cause a change of the supercharge from the tree-level supercharge.}

The second supercharge has the commutation relations
\begin{subequations}
\begin{align}
    \left[\mathfrak{Q'}_+,\mathfrak{L'}_+\right]\ket{(q,n)(k-q,m-n)}=\hspace{-10em}&\cr
    =~&\Bigl[\delta_{n\neq{0}}(q-n+2)c^+_{q,n-1;k,m-1}
    +    \cr&
    +\delta_{m-n\neq{0}}(k-q-m+n+2)c^+_{q,n;k,m-1}
    +    \cr&
    -\delta_{m\neq{0}}(k-m+3)c^+_{q,n;k,m}\Bigr]\ket{(k+1,m-1)}
\\
    \left[\mathfrak{Q'}_+,\mathfrak{L'}_-\right]\ket{(q,n)(k-q,m-n)}=\hspace{-10em}&\cr
    =~&\Bigl[\delta_{q-n+1\neq{0}}(n+1)c^+_{q,n+1;k,m+1}
    +    \cr&
    +\delta_{k-q-m+n+1\neq{0}}(m-n+1)c^+_{q,n;k,m+1}
    +   \cr&
    -\delta_{k-m+2\neq{0}}(m+1)c^+_{q,n;k,m}\Bigr]\ket{(k+1,m+1)}
\\
    \left[\mathfrak{Q'}_+,\mathfrak{J'}_{-+}\right]\ket{(q,n)(k-q,m-n)}=\hspace{-10em}&\cr
    =~&\Bigl[(q-n+2)c^+_{q+1,n;k+1,m}
    +(k-q-m+n+2)c^+_{q,n;k+1,m}
    +   \cr&
    -(k-m+3)c^+_{q,n;k,m}\Bigr]\ket{(k+2,m)}
\\
    \left[\mathfrak{Q'}_+,\mathfrak{J'}_{+-}\right]\ket{(q,n)(k-q,m-n)}=\hspace{-10em}&\cr
    =~&\Bigl[\delta_{q-n+1\neq{0}}qc^+_{q-1,n;k-1,m}
    +\delta_{k-q-m+n+1\neq{0}}(k-q)c^+_{q,n;k-1,m}
    +   \cr&
    -\delta_{k-m+2\neq{0}}(k+1)c^+_{q,n;k,m}\Bigr]\ket{(k,m)}
\\
    \left[\mathfrak{Q'}_+,\mathfrak{J'}_{--}\right]\ket{(q,n)(k-q,m-n)}=\hspace{-10em}&\cr
    =~&\Bigl[(n+1)c^+_{q+1,n+1;k+1,m+1}
    +(m-n+1)c^+_{q,n;k+1,m+1}
    +   \cr&
    -(m+1)c^+_{q,n;k,m}\Bigr]\ket{(k+2,m+1)}
\\
    \left[\mathfrak{Q'}_+,\mathfrak{J'}_{++}\right]\ket{(q,n)(k-q,m-n)}=\hspace{-10em}&\cr
    =~&\Bigl[\delta_{n\neq{0}}q\,c^+_{q-1,n-1;k-1,m-1}
    +\delta_{m-n\neq{0}}(k-q)c^+_{q,n;k-1,m-1}
    +   \cr&
    -\delta_{m\neq{0}}(k+1)c^+_{q,n;k,m}\Bigr]\ket{(k,m-1)}~.
\end{align}
\end{subequations}
All the commutators above must vanish up to a gauge transformation. The unique solution is
\begin{equation}
    c^+_{q,n;k,m}
    =\frac{(k+2)}{(q+1)(k-q+1)}\frac{\binom{q+1}{n}\binom{k-q+1}{m-n}}{\binom{k+2}{m}}\,a_+~,
\end{equation}
at which the commutators take the value
\begin{subequations}
\begin{align}
    &\left[\mathfrak{Q'}_+,\mathfrak{L'}_{+}\right]\ket{(q,n)(k-q,m-n)}=0
\\
    &\left[\mathfrak{Q'}_+,\mathfrak{L'}_{-}\right]\ket{(q,n)(k-q,m-n)}=0
\\
    &\left[\mathfrak{Q'}_+,\mathfrak{J'}_{-+}\right]\ket{(q,n)(k-q,m-n)}=0
\\
    &\left[\mathfrak{Q'}_+,\mathfrak{J'}_{+-}\right]\ket{(q,n)(k-q,m-n)}=-a_+\left(\delta_{q=0}+\delta_{q=k}\right)\ket{(k,m+1)}
\\
    &\left[\mathfrak{Q'}_+,\mathfrak{J'}_{--}\right]\ket{(q,n)(k-q,m-n)}=0
\\
    &\left[\mathfrak{Q'}_+,\mathfrak{J'}_{++}\right]\ket{(q,n)(k-q,m-n)}=0~.
\end{align}
\end{subequations}
Summarizing the above we find
\begin{align}
    \mathfrak{Q'}_+\ket{(q,n)(k,m)}
    &=a_+\frac{(k+q+2)}{(q+1)(k+1)}\frac{\binom{q+1}{n}\binom{k+1}{m}}{\binom{k+q+2}{m+n}}\ket{(k+q+1,m+n)}
\\
    \mathfrak{Q'}_-\ket{(k,m)}=&a_-\sum_{q=0}^{k-1}\sum_{\substack{n=\\\max(0,m+q-k)}}^{\min(q+1,m)}\ket{(q,n)(k-q-1,m-n)}~.
\end{align}
The 1-loop spin-chain Hamiltonian ($\dD_2$) can be calculated from the commutator
\begin{equation}\label{com0}
    \dD=2\bigl\{\mathfrak{Q'}_+,\mathfrak{Q'}_-\bigr\}~.
\end{equation}
Acting on a two-spin state. First we calculate
\begin{align}
    \mathfrak{Q'_-}\ket{(q,n)(k,m)}
    =&
    +a_-\sum_{s=0}^{q-1}\sum_{\substack{l=\\\max(0,n+s-q)}}^{\min(s+1,n)}\ket{(s,l)(q-s-1,n-l)(k,m)}+\cr
    &-a_-\sum_{s=0}^{k-1}\sum_{\substack{l=\\\max(0,m+s-k)}}^{\min(s+1,m)}\ket{(q,n)(s,l)(k-s-1,m-l)}~.
\end{align}
In the spin chain limit $\mathfrak{Q'}_+$ acts only on adjacent partons, thus we find
\begin{align}\label{com1}
    \mathfrak{Q'}_+&\mathfrak{Q'}_-\ket{(q,n)(k,m)}=a_+a_-\Biggl\{\cr
    &+\frac12\sum_{s=0}^{q-1}\sum_{\substack{l=\\\max(0,n+s-q)}}^{\min(s+1,n)}\hspace{-1em}
    \frac{(q+1)}{(s+1)(q-s)}\frac{\binom{s+1}{l}\binom{q-s}{n-l}}{\binom{q+1}{n}}\ket{(q,n)(k,m)}+\cr
    &-\sum_{s=0}^{q-1}\sum_{\substack{l=\\\max(0,n+s-q)}}^{\min(s+1,n)}\hspace{-.5em}
    \frac{(k+q-s+1)}{(k+1)(q-s)}\frac{\binom{q-s}{n-l}\binom{k+1}{m}}{\binom{k+q-s+1}{m+n-l}}\ket{(s,l)(k+q-s,m+n-l)}+\cr
    &-\sum_{s=q+1}^{k+q}\sum_{\substack{l=\\\max(n,n+m+s-1-k-q)}}^{\min(n+s-q,m+n)}\hspace{-1.8em}
    \frac{(s+1)}{(q+1)(s-q)}\frac{\binom{q+1}{n}\binom{s-q}{l-n}}{\binom{s+1}{l}}\ket{(s,l)(k+q-s,m+n-l)}+\cr
    &+\frac12\sum_{s=0}^{k-1}\sum_{\substack{l=\\\max(0,m+s-k)}}^{\min(s+1,m)}\hspace{-1em}
    \frac{(k+1)}{(s+1)(k-s)}\frac{\binom{s+1}{l}\binom{k-s}{m-l}}{\binom{k+1}{m}}\ket{(q,n)(k,m)}
    \Biggr\}~.
\end{align}
The factor of $\frac12$ comes due to symmetrization of the operator as explained in \cite{Beisert:2004ry} for the $\su(1,1)$ case.

Acting with the commutator \eqref{com0} on a single spin state we find
\begin{multline}\label{com2}
    \mathfrak{Q'_-}\mathfrak{Q'_+}\ket{(q,n)(k,m)}
    =a_-a_+
    \sum_{s=0}^{k+q}\sum_{\substack{l=\\\max(0,m+n+s-k-q-1)}}^{\min(s+1,m+n)}\hspace{-1em}
    \times\\\times
    \frac{(k+q+2)}{(q+1)(k+1)}\frac{\binom{q+1}{n}\binom{k+1}{m}}{\binom{k+q+2}{m+n}}\ket{(s,l)(k+q-s,m+n-l)}~.
\end{multline}
Applying \eqref{com1}, \eqref{com2} to \eqref{com0} we find (setting $2a_+a_-=1$)
\begin{align}
    \dD_2\ket{(q,n)(k,m)}=\hspace{-3em}&\hspace{3em}
    \Bigl[h(q)+h(k)\Bigr]\ket{(q,n)(k,m)}+\cr
    &+\Bigg[-\sum_{s=0}^{q-1}\sum_{\substack{l=\\\max(0,n+s-q)}}^{\min(s+1,n)}
    \frac{(k+q-s+1)}{(k+1)(q-s)}\frac{\binom{q-s}{n-l}\binom{k+1}{m}}{\binom{k+q-s+1}{m+n-l}}+\cr
    &\quad-\sum_{s=q+1}^{k+q}\sum_{\substack{l=\\\max(n,n+m+s-1-k-q)}}^{\min(n+s-q,m+n)}
    \frac{(s+1)}{(q+1)(s-q)}\frac{\binom{q+1}{n}\binom{s-q}{l-n}}{\binom{s+1}{l}}+\cr
    &\quad+\sum_{s=0}^{k+q}\sum_{\substack{l=\\\max(0,m+n+s-k-q-1)}}^{\min(s+1,m+n)}
    \frac{(k+q+2)}{(q+1)(k+1)}\frac{\binom{q+1}{n}\binom{k+1}{m}}{\binom{k+q+2}{m+n}}
    \Bigg]\cr&\qquad\times\ket{(s,l)(k+q-s,m+n-l)}\,
\end{align}
the lifting of the spin chain Hamiltonian to finite N is done using \eqref{spin-finite1} and \eqref{spin-finite2}

\end{document}